\documentclass[12pt]{article}

\usepackage{bbold}
\usepackage{latexsym}
\usepackage{slashed}
\usepackage{graphics}

\declareslashed{}{/}{.1}{0}{E}

\def\ve{\varepsilon}
\def\ss{\scriptstyle}
\def\sss{\scriptscriptstyle}

\newcommand{\sy}[1]{(\hspace{-.01mm}\underline{#1})}
\newcommand{\syt}[1]{(\hspace{-.01mm}\underline{#1})^{\!\rm T}}
\newcommand{\sytt}[1]{(\hspace{-.01mm}\underline{#1})^{\!\rm T\!T}}
\newcommand{\syg}[1]{(\hspace{-.01mm}\underline{#1})^{\rm \gamma}}
\newcommand{\sygt}[1]{(\hspace{-.01mm}\underline{#1})^{\rm \gamma T}}

\newcommand{\DB}[1]{\Delta^{\!(#1)}}
\newcommand{\DF}[1]{{\cal D}^{(#1)}}
\newcommand{\lDF}[1]{\overleftarrow{\cal D}^{(#1)}}

\newcommand{\Dp}[1]{D^{(#1)}}
\renewcommand{\S}[1]{S^{(#1/2)}}

\def\TT{{\rm TT}}

\def\chib{\overline\chi}
\def\psib{\overline\psi\hspace{-2.6mm}\phantom{\psi}}

\def\ve{\varepsilon}
\def\D{{\cal D}}

\def\r{\rho}
\def\s{\sigma}

\def\Sl#1{\slashed{#1}}
\def\sl#1{\slashed{#1}}
\def\wh#1{\widehat{#1}}
\def\wt#1{\widetilde{#1}}

\def\ol#1{\overline{#1}} 
\def\sss{\scriptscriptstyle}

\def\d{\partial}
\def\m{\mu}
\def\n{\nu}

\def\e{\epsilon}

\def\be{\begin{equation}}
\def\ee{\end{equation}}
\def\beq{\begin{equation}}
\def\eeq{\end{equation}}
\def\bea{\begin{eqnarray}}
\def\eea{\end{eqnarray}} 
\def\beqa{\begin{equation}\begin{array}{l}}
\def\eeqa{\end{array}\end{equation}}

\def\eqn#1{(\ref{#1})}

\def\eqref#1{eq.~(\ref{eq:#1})}

\def\a{\alpha}
 \def\G{{\it\Gamma}} \def\g{\gamma}

\def\L{{\it\Lambda}}

\def\nn{\nonumber}

\begin{document}

\thispagestyle{empty}
\begin{flushright}
\framebox{\small BRX-TH~486
}\\
\end{flushright}

\vspace{.8cm}
\setcounter{footnote}{0}
\begin{center}
{\Large{\bf 
Partial~Masslessness~of~Higher~Spins~in~(A)dS}
    }\\[10mm]

{\sc S. Deser
and A. Waldron
\\[6mm]}

{\em\small  
Physics Department, Brandeis University, Waltham,
MA 02454, 
USA\\ {\tt deser,wally@brandeis.edu}}\\[5mm]

\bigskip

\bigskip

{\sc Abstract}\\
\end{center}

{\small
\begin{quote}

Massive spin $s\geq3/2$ fields 
can become partially massless in cosmological 
backgrounds. In the plane spanned by $m^2$ and $\L$, there
are lines where new gauge invariances
permit intermediate sets of higher helicities, rather
than the usual flat space extremes of
all $2s+1$ massive or just 2 massless 
helicities.
These gauge lines divide the $(m^2,\Lambda)$ plane
into unitarily allowed or forbidden intermediate regions
where all $2s+1$ massive helicities propagate but lower helicity states 
can have negative norms. We derive these consequences 
for $s=3/2,2$ by studying both their canonical (anti)commutators
and the transmutation of
massive constraints to partially massless Bianchi identities.
For $s=2$, a Hamiltonian analysis exhibits the absence of
zero helicity modes
in the partially massless
sector. For $s=5/2,3$ we derive
Bianchi identities and their accompanying  gauge invariances for the
various partially massless theories with propagating helicities
$(\pm5/2,\pm3/2)$ and $(\pm3,\pm2)$, $(\pm3,\pm2,\pm1)$,
respectively. Of these, only the $s=3$ models are unitary.
To these ends, we also provide the half integer
generalization of the integer spin wave operators of Lichnerowicz.
Partial masslessness applies to all higher spins in (A)dS 
as seen by their degree of freedom counts.
Finally a derivation of massive $d=4$ constraints by dimensional reduction
{}from their $d=5$ massless
Bianchi identity ancestors is  given.

\bigskip

\bigskip

\end{quote}
}


\newpage





\section{Introduction}

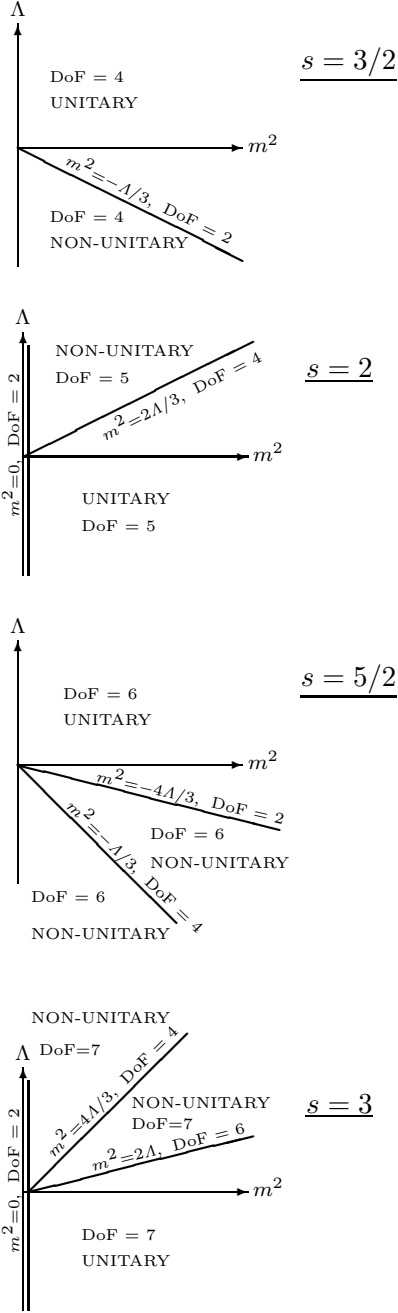
\begin{figure}

\bea
\begin{picture}(300,100)(-100,5)
\put(0,100){$\ss \Lambda$}
\put(90,48.5){$\ss m^2$}


\put(20,44){\rotatebox{334}{$\sss m^2=-\L/3 ,\;\;\mbox{\tiny DoF = 2}$}}
\put(15,75){\tiny DoF = 4}
\put(15,65){\tiny UNITARY}
\put(15,22){\tiny DoF = 4}
\put(15,12){\tiny NON-UNITARY}
\put(110,80){\small \underline{$s=3/2$}}



\put(3,50){\vector(1,0){85}}
\put(3,5){\vector(0,1){92}}

\thicklines

\put(3,50){\line(2,-1){85}}




\end{picture}
\nn
\eea
\vspace{-.8cm}
\bea
\begin{picture}(300,100)(18,5)

\put(120,100){$\ss \Lambda$}
\put(210,48.5){$\ss m^2$}

\put(151,56){\rotatebox{27}{$\sss m^2=2\L/3 ,\;\;\mbox{\tiny DoF = 4}$}}
\put(115,27){\rotatebox{90}{$\sss m^2=0 ,\;\;\mbox{\tiny DoF = 2}$}}
\put(135,88){\tiny NON-UNITARY}
\put(135,78){\tiny DoF = 5}
\put(145,22){\tiny DoF = 5}
\put(145,32){\tiny UNITARY}

\put(230,80){\small \underline{$s=2$}}

\put(123,50){\vector(1,0){85}}
\put(123,5){\vector(0,1){92}}

\thicklines

\put(125,5){\line(0,1){87}}
\put(123,50){\line(2,1){87}}

\end{picture}
\nn
\eea
\vspace{-.8cm}
\bea
\begin{picture}(300,120)(-100,-15)
\put(0,100){$\ss \Lambda$}
\put(90,48.5){$\ss m^2$}

\put(32,44){\rotatebox{347}{$\sss m^2=-4\L/3 ,\;\;\mbox{\tiny DoF = 2}$}}
\put(20,34){\rotatebox{316}{$\sss m^2=-\L/3 ,\;\;\mbox{\tiny DoF = 4}$}}
\put(20,75){\tiny DoF = 6}
\put(20,65){\tiny UNITARY}
\put(53,22){\tiny DoF = 6}
\put(53,11){\tiny NON-UNITARY}
\put(8,-2){\tiny DoF = 6}
\put(8,-16){\tiny NON-UNITARY}

\put(110,80){\small \underline{$s=5/2$}}

\put(3,50){\vector(1,0){85}}
\put(3,5){\vector(0,1){92}}

\thicklines

\put(3,50){\line(4,-1){99}}
\put(3,50){\line(1,-1){60}}

\end{picture}
\nn
\eea
\vspace{-.1cm}
\bea
\begin{picture}(300,100)(138,10)

\put(240,100){$\ss \Lambda$}
\put(330,48.5){$\ss m^2$}

\put(235,27){\rotatebox{90}{$\sss m^2=0 ,\;\;\mbox{\tiny DoF = 2}$}}
\put(250,63){\rotatebox{45}{$\sss m^2=4\L/3 ,\;\;\mbox{\tiny DoF = 4}$}}
\put(268,58){\rotatebox{14}{$\sss m^2=2\L ,\;\;\mbox{\tiny DoF = 6}$}}

\put(265,32){\tiny DoF = 7}
\put(265,22){\tiny UNITARY}
\put(284,82){\tiny NON-UNITARY}
\put(284,74){\tiny DoF=7}
\put(246,114){\tiny NON-UNITARY}
\put(249,102){\tiny DoF=7}

\put(350,80){\small \underline{$s=3$}}

\put(243,50){\vector(1,0){85}}
\put(243,5){\vector(0,1){92}}

\thicklines

\put(245,5){\line(0,1){87}}
\put(245,50){\line(4,1){85}}
\put(245,50){\line(1,1){60}}

\end{picture}
\nn
\eea

\label{regions}
\caption{Respective phase diagrams for spins $s=3/2,2,5/2,3$ showing
partially massless gauge lines and unitarily allowed and forbidden regions.}
\vspace{1cm}
\end{figure}

The $(m^2,\Lambda)$ mass-cosmological constant plane for spin
$s=2$ is 
unexpectedly rich~\cite{Deser:1983tm,Higuchi:1987py,Deser:2001pe}: 
For generic $(m^2,\Lambda)$, 
the theory describes 5 massive degrees of freedom. However, the 
field equations,
having two open indices, are susceptible to both single and double
divergence Bianchi identities each signaling the onset of gauge invariances.
This is indeed realized when $m^2=0$ and $m^2=2\Lambda/3$,
respectively. The former corresponds to linearized 
cosmological Einstein gravity and the latter to the partially massless 
de Sitter (dS) theory of~\cite{Deser:1983tm}.
A peculiarity of (Anti) de Sitter ((A)dS) spaces 
is that gauge invariance does not necessarily
imply null propagation, which in fact occurs only for the
$m^2=2\Lambda/3$ theory~\cite{Deser:1983tm}.
Furthermore, these two gauge invariant theories lie on the boundary of
two physically distinct regions (see Figure~\ref{regions}). 
As demonstrated in~\cite{Higuchi:1987py}, 
in the dS region with masses bounded by $0<m^2<2\Lambda/3$,
the massive $s=2$ spectrum is not unitary. This is an example of a
generic difficulty encountered by higher spin theories in
backgrounds; it was first observed in a study of
canonical anticommutators for charged massive $s=3/2$ fields  
in electromagnetic backgrounds~\cite{Johnson:1961vt} and later shown to imply
acausal propagation~\cite{Velo:1969bt}.

The above $s=2$ behaviors are common spins $s\geq 3/2$ in
(A)dS, as seen by studying possible Bianchi identities.
The number of massive higher spin covariant field components exceeds
the $2s+1$ propagating degrees of freedom (PDoF) count. The correct
count is
imposed by constraints constructed from divergences
of each open index of the field equations,  $D^{\m_1}\cdots
D^{\m_n}\,{\cal G}_{\m_1\ldots\m_n\m_{n+1}\ldots}$ $+$ $\cdots=0$. 
In flat space, all these
constraints hold identically when the mass $m$ vanishes and are 
encapsulated by a single Bianchi identity 
$\d^{\m_1}{\cal G}_{\m_1\m_2\ldots}+\cdots=0$ (for $s>2$ the terms
$+\cdots$ ensure (gamma-)tracelessness of the remaining open indices)
reflecting the gauge invariance of the massless theory~\cite{deWit:1980pe}.
Once a cosmological constant is present, this degeneracy is broken
and for special tunings of $m^2$ to $\L$, intermediate situations are 
possible where some of the lower divergence constraints remain and 
all the higher ones are replaced by a single Bianchi 
identity~\cite{Deser:2001pe}. This new identity 
entails a gauge invariance of the theory.
As a consequence, higher spins in
(A)dS spaces can be partially massless: not only are there strictly
massive theories with all $2s+1$ 
helicities  and strictly massless ones with only 
helicities $\pm s$, but also intermediate theories of helicities
$(\pm s,\pm(s-1),\ldots,\pm(s-t))$~($t<s$)~\footnote{An additional subtlety
for $s\geq5/2$ is that auxiliary fields are necessary (over and
above the field content of the strictly massless models), but at least
in the massless limit, these can be shown to decouple.}.

The new Bianchi identities appear along lines in the $(m^2,\Lambda)$
plane and are also manifested in the canonical (anti)commutators of
the fields.  These follow directly from 
the covariant transverse-traceless decomposition of the on-shell
fields\footnote{Their restriction to equal time commutators agrees
with the detailed $s=2$ Dirac analysis
of~\cite{Higuchi:1987py}.}~\cite{Deser:2001pe}.   
The gauge lines in the $(m^2,\Lambda)$ plane
correspond to poles in the lower helicity terms of (anti)commutators,
the usual indication that a theory enjoys a gauge invariance at special
values of parameters. Therefore, lower helicity (anti)commutators 
flip sign across these gauge lines. The
intermediate
regions correspond to (unitarily allowed or forbidden) massive
theories; the one including the massive Minkowski theory is always
unitary.
The norms of those helicities that are removed by a gauge 
invariance flip sign across the corresponding line.
This implies that the partially massless theories of the remaining
helicities are unitary only when, starting from
the
unitary Minkowski region, the line in the $(m^2,\Lambda)$ half-plane 
removing the lowest helicity
state(s)
can be traversed first and subsequent lines are encountered in order,
ending on the strictly massless helicity $\pm s$ gauge line. 
The latter is guaranteed to be unitary since it removes all 
dangerous lower helicities. We show explicitly that this condition is
fulfilled for $s=2,3$ but not for the novel $s=5/2$ partially
massless gauge line. We will also demonstrate, by a simple counting argument,
that partial masslessness is enjoyed by all higher spins in (A)dS.

In Section~\ref{begin} we introduce
massive
spins $s\leq2$ and their gauge invariances. We also summarize 
the higher integer spin wave
operators of Lichnerowicz~\cite{Lichnerowicz:1961} 
and provide their generalizations to 
half integer spins. In Section~\ref{keepgoing}, we study canonical
(anti)commutators for $s\leq2$ and show how non-unitary regions
arise. In Section~\ref{whale}, we examine the partially massless
$m^2=2\L/3$,  $s=2$ theory, demonstrating by Hamiltonian analysis that it 
describes propagating helicities $(\pm2,\pm1)$ only. Section~\ref{dolphin}
deals with all spins: We first present a counting argument
indicating that partial masslessness occurs for all higher spins in (A)dS.
Then we display explicit new Bianchi identities and
partially massless gauge invariances for $s=5/2,3$.
The Appendix extends the  
dimensional reduction derivation of massive higher spin field equations
in flat space from their massless dimension $d=5$ 
predecessors~\cite{Aragone:1987yx},
to obtain massive constraints
{}from 
the $d=5$ Bianchi identity for $s=5/2,3$ examples.
Some future directions and open problems are discussed in the Conclusion.
A brief version of some of our results was given in~\cite{Deser:2001pe}.

\section{Field Equations and Identities in Constant Curvature Spaces}

\label{begin}

The Riemann tensor in constant curvature spaces is
\be
R_{\m\n\r\s}=-\frac{2\L}{3}\,g_{\m[\r}g_{\s]\n}\ ; 
\ee
the cosmological constant $\L$ is positive in dS 
and negative in AdS. The actions of commutators of covariant 
derivatives are summarized by the vector-spinor example\footnote{Our 
metric is ``mostly plus'',
Dirac matrices are ``mostly hermitean'' and the Dirac conjugate is
$\ol \psi\equiv \psi^\dagger i\g^0$. 
We denote (anti)symmetrization with unit weight by round
(resp. square) brackets. Antisymmetrized products of Dirac matrices
are given by $\g^{\m_1\ldots\m_n}\equiv\g^{[\m_1}\cdots\g^{\m_n]}$.}
\be
[D_\m,D_\n]\,\psi_\r=\frac{2\L}{3}\,g_{\r[\m}\psi_{\n]}
+\frac{\L}{6}\,\g_{\m\n}\psi_\r\, .
\ee

The actions and field equations for massive spins $0\leq s\leq2$ 
in constant curvature backgrounds\footnote{For $s=2$ in generic
gravitational backgrounds, the minimally coupled
Pauli--Fierz action 
does not yield field equations with the correct $2s+1=5$ 
PDoF~\cite{Boulware:1972}. Recently~\cite{Buchbinder:2000ar}, $s=2$ 
actions have been constructed with the correct PDoF count in
background Einstein spaces 
($R_{\m\n}=\L g_{\m\n}$).} are
\be
\begin{array}{cc}
{\cal L}^{(0)}=\frac{1}{2}\,\sqrt{-g}\;\phi\;{\cal G}\ ,&\\
&{\cal L}^{(1/2)}=-\sqrt{-g}\; \psib\,{\cal R}\ ,\\
{\cal L}^{(1)}=\frac{1}{2}\,\sqrt{-g}\;\phi^\m\;{\cal G}_\m\ ,&\\
&{\cal L}^{(3/2)}=-\sqrt{-g}\; \psib^\m\,{\cal R}_\m\ ,\nn\\
\quad\;\;\;{\cal L}^{(2)}=\frac{1}{2}\,\sqrt{-g}\;\phi^{\m\n}\;
{\cal G}_{\m\n}\ ,\qquad&
\end{array}
\label{ramit}
\ee
\bea
{\cal G}&=&(\DB{0}-m^2)\,\phi\, ,\\
{\cal R}&=&(\DF{1/2}+m)\,\psi\, ,\\
{\cal G}_\m&=&(\DB1-m^2)\,\phi_\m-D_\m D.\phi\, ,\label{Proca}\\
{\cal R}_\m&\equiv&\g_{\m\n\r}\D^\n\psi^\r
=(\DF{3/2}-\Sl D)\,\psi_\m-m\,\g_{\m\n}\psi^\n\, ,\label{RS}\\
{\cal G}_{\m\n}&\equiv&
(\DB2-m^2+\L)\,(\phi_{\m\n}-g_{\m\n}\phi_\r{}^\r)+\L\phi_{\m\n}\nn\\&+&
\!D_{(\m}D_{\n)}\,\phi_\r{}^\r-2D_{(\m}D.\phi_{\n)}
+g_{\m\n}D.D.\phi\, .\label{ES}
\eea
The operators $\DB{n}$ are the wave operators of~\cite{Lichnerowicz:1961}
\bea
\DB3\,\phi_{\m\n\r}&\equiv&D^2\,\phi_{\m\n\r}-5\L\,\phi_{\m\n\r}+
2\L\,g_{(\m\n}\phi_{\r)\s}{}^\s\,\ , \\
\DB2\,\phi_{\m\n}&\equiv&D^2\phi_{\m\n}
-\frac{8\,\L}{3}\,\Big(\phi_{\m\n}-\frac{1}{4}\,g_{\m\n}\,
\phi_\r{}^\r\Big) \ ,\\
\DB1\,\phi_\m&\equiv&(D^2-\L)\,\phi_\m \ ,\\
\DB0\,\phi&\equiv&D^2\phi \ ,
\eea
whose introduction is justified by the following identities,
\be
\begin{array}{cc}
\DB3\, D_{(\m} \phi_{\n\r)} =
D_{(\m}\DB2\,\phi_{\n\r)}\, ,&
D^\m \DB3\,\phi_{\m\n\r}=\DB2\,D.\phi_{\n\r}\, ,\\ \\ 
\DB3\,g_{(\m\n}\phi_{\r)}=g_{(\m\n}\DB1 \,\phi_{\r)}\, ,&
g^{\m\n}\DB3 \phi_{\m\n\r}=\DB1\,\phi^\s{}_{\s\r}\, ,\\ \\
\DB2\, D_{(\m} \phi_{\n)} =
D_{(\m}\DB1\,\phi_{\n)}\, ,&
D^\m \DB2\,\phi_{\m\n}=\DB1\,D.\phi_\n\, ,\\ \\ 
\DB2\,g_{\m\n}\phi=g_{\m\n}\DB0 \,\phi\, ,&
g^{\m\n}\DB2 \phi_{\m\n}=\DB0\,\phi^\r{}_\r\, ,\\ \\
\DB1 D_\m\phi= D_\m \DB0 \phi \, ,&
D^\m \DB1\,\phi_\m = \DB0\,D.\phi \, .\\ 
\end{array}
\label{ids}
\ee
Here $\phi_{\m\n\ldots}$ is a symmetric tensor. 
The half integer spin generalizations
$\DF{n/2}$ of these operators are given by~\cite{Deser:2001de}
\bea
\hspace{-.5cm}
\DF{5/2}\,\psi_{\m\n}&\!\equiv\!&
3\,\Sl D\,\psi_{\m\n}-2\,D_{(\m}\g.\psi_{\n)}
+2\,\g_{(\m}\,(\Sl D\,\g.\psi_{\n)}-D.\psi_{\n)})\ ,
\\
\DF{3/2}\,\psi_\m&\!\equiv\!&
2\sl D\,\psi_\m-D_\m\g.\psi+
\g_\m(\sl D\g.\psi-D.\psi)=\g_{\m\n\r}D^\n \psi^\r+\sl D\psi_\m\ ,\quad \\
\DF{1/2}\,\psi&\!\equiv\!&\sl D\,\psi\ , 
\eea
with $\psi_{\m\n}$ a symmetric spinor-tensor,
and satisfy analogous identities  
\be
\begin{array}{cc}
\DF{5/2} \,D_{(\m}\,\psi_{\n)}=D_{(\m}\, \DF{3/2}\,\psi_{\n)}\, , &
 D^\m\,\DF{5/2}\,\psi_{\m\n}=\DF{3/2}\,D.\psi_\n \, ,\\ \\
\DF{5/2} \,\g_{(\m}\,\psi_{\n)}=\g_{(\m}\, \DF{3/2}\,\psi_{\n)}\, ,&
\g^\m\,\DF{5/2}\,\psi_{\m\n}=\DF{3/2}\,\g.\psi_\n \, ,\\ \\
\DF{3/2} \,D_\m\,\psi=D_\m\, \DF{1/2}\,\psi\, , &
 D^\m\,\DF{3/2}\,\psi_\m=\DF{1/2}\,D.\psi \, ,\\ \\
\DF{3/2} \,\g_\m\,\psi=\g_\m\, \DF{1/2}\,\psi\, ,&
\g^\m\,\DF{3/2}\,\psi_\m=\DF{1/2}\,\g.\psi \, .
\end{array}
\ee
For completeness we include the familiar Weitzenbock identity,
\be
\Big(\DF{1/2}\Big)^2\,\psi=(D^2-\L)\,\psi\, .
\ee
Notice that we have written the massive  Rarita--Schwinger field
equation~\eqn{RS} in terms of the $s=3/2$ operator $\DF{3/2}$ with
explicit mass term as well as in a more compact form involving the operator
\be
\D_\m\equiv D_\m+\frac{m}{2}\,\g_\m\, ,\qquad
[\D_\m,\D_\n]=[D_\m,D_\n]+(m^2/2)\,\g_{\m\n}\, ,
\ee
encountered in cosmological supergravity~\cite{Townsend:1977qa}.

For $s\geq1$ there are more relativistic field components than
PDoF. As usual, the correct PDoF count is obtained by studying
the constraints implied by divergences and (gamma-)traces of the
field equations. Less usual, for special lines in the $(m^2,\Lambda)$
plane,  these constraints are satisfied identically and become 
Bianchi identities associated with gauge invariances. Explicitly, 
the divergence of the field equations~\eqn{Proca}-\eqn{ES} read 
\be
\begin{array}{cc}
D.{\cal G}=-m^2\,D.\phi\, ,&\\
&\D.{\cal R}=-\frac{1}{2}\,(3m^2+\L)\,\g.\psi\, ,\\
D.{\cal G}_\n=-m^2\,(D.\phi_\n-D_\n\phi)&
\end{array}
\label{const}
\ee
and are constraints for generic values of the parameters 
$m^2$ and $\L$. For $s=1,2$, the value $m^2=0$ yields Bianchi identities
and their associated (``electromagnetic'' and ``general coordinate'') 
gauge invariances
\be
\delta \phi_\m=D_\m \xi \qquad\mbox{and}\qquad 
\delta \phi_{\m\n}=D_{(\m}\xi_{\n)}\, ,
\ee
respectively. Both $m^2=0$ theories are strictly massless: they
propagate with two physical helicity states.
For $s=3/2$, the sole gauge invariance 
\be
\delta \psi_\m=\D_\m \varepsilon
=D_\m\varepsilon+\frac{1}{2}\,\sqrt{-\L/3}\,\g_\m\,\varepsilon \, ,
\ee
is inherited from 
cosmological supergravity and occurs for $m^2=-\L/3$ in AdS.
This (rather than the $m=0$ model) 
is the strictly massless helicity $\pm3/2$ theory.

For $s=2$ there is a new effect: the field equation has two open
indices and thereby also admits a double divergence Bianchi identity.
The double divergence constraint
\be
D.D.{\cal G}+\frac{1}{2}\,m^2\,{\cal
G}_\r{}^\r=\frac{1}{2}\,m^2\,(3m^2-2\L)\,\phi_\rho{}^\rho\, ,
\label{sticky}
\ee
becomes a Bianchi identity not only along the strictly
massless line $m^2=0$,
but also along the dS gauge line $m^2=2\L/3$ corresponding to the Weyl-like
theory of~\cite{Deser:1983tm}. The relevant gauge invariance 
\be
\delta \phi_{\m\n}=\Big(D_{(\m}D_{\n)}+\frac{\L}{3}\,g_{\m\n}\Big)\;\xi\, ,
\ee
may be employed to show that this partially massless
model propagates with helicities
$(\pm2,\pm1)$ on the null cone in dS. We  discuss this theory 
further in Section~\ref{whale}.

To summarize, the $(m^2,\Lambda)$ half-plane offers no surprises for
spins $s\leq1$,  the usual null propagating massless theories
inhabit the line $m^2=0$ and all other points $m^2\geq0$ describe $2s+1$
massive DoF. For $s=3/2$, the massless line is $m^2=-\L/3$ and
bisects the
$(m^2,\Lambda)$ half-plane as depicted in Figure~\ref{regions}. For
$s=2$, the massless theory again lies on the axis $m^2=0$ and a new
gauge invariance emerges at $m^2=2\L/3$ which also divides the half-plane
into two distinct physical regions.
  
\section{\hspace{-.5cm}(Anti)commutators and Nonunitary Regions}

\label{keepgoing}

Our analysis is rather simple and harks back
to the original inconsistency of the local field theory of charged
spin~3/2 particles~\cite{Johnson:1961vt}. Generically, given (anti)commutation
relations $\{\psi,\psi^\dagger\}=\epsilon=[a,a^\dagger]=-i[x,\dot x]$
(with $a=(x+i\dot x)/\surd{2}$) {\it and} a vacuum\footnote{It has
recently been suggested that the non-perturbative definition of
quantum gravity in dS be reexamined~\cite{Witten:2001}; 
in particular the
definition of the vacuum requires careful consideration. Our local
quantum field theoretic computation ignores such subtleties.}
$\psi\,|0\rangle=0=a\,|0\rangle$, positivity of norms requires
$\epsilon>0$. For quantum field theories, exactly the same criterion
can be applied, but now in a distributional sense.

The local canonical (anti)commutators for quantum fields with 
$s\leq 2$ in cosmological spaces 
were presented long ago~\cite{Lichnerowicz:1961}. Let us summarize
those results: For $s=0,1/2$ one has
\bea
[\phi(x),\phi(x')]&=&i \,\Dp0(x,x';m^2)\, ,\label{nil}\\
\{\psi(x),\ol \psi(x')\}&=& i\, \S1(x,x';m)\, .\label{halb}
\eea
The distributions on the right hand sides of~\eqn{nil} and~\eqn{halb}
are general coordinate/local Lorentz biscalars and 
bivectors~\cite{DeWitt:1960fc}, 
respectively, and share the symmetry properties of the
(anti)commutators on the left hand sides
\be
\Dp0(x,x')=-\Dp0(x,x')\; ,\qquad
\S1(x,x') \g^0 = - \g^0 \Big[\S1(x',x)\Big]^\dagger \, . 
\ee
We often drop the the labels ``$m^2$'' and ``$m$'' used to indicate
that these distributions satisfy the field equations of the onshell
quantum fields $\phi(x)$ and $\psi(x)$
\be
(\DB0_x-m^2)\,\Dp0(x,x';m^2)=0=(\DB0_{x'}-m^2)\,\Dp0(x,x';m^2)\; ,
\ee
\be
(\DF{1/2}+m)\,\S1(x,x';m)=0=\S1(x,x';m)\,(-\lDF{1/2}_{x'}+m)\, .
\ee 
Here the backwards arrow over an operator ${\cal O}$ is defined by
\be
\ol \psi \,\overleftarrow{\cal O}\equiv-\,\ol{{\cal O}\psi}\, . 
\ee
The boundary conditions for these distributions are
$$
\frac{d}{dx'{}^0}\Dp0(x,x')\Big|_{x^0=x'{}^0}=
\frac{1}{\sqrt{-g}}\,\delta^3(\vec x-\vec x\,')\;,\quad\
\Dp0(x,x')\Big|_{x^0=x'{}^0}=0\, ,
$$
\be\label{ing}\ee
\vspace{-1cm}
$$
\S1(x,x')\Big|_{x^0=x'{}^0}=
\frac{\g^0}{\sqrt{-g}}\,\delta^3(\vec x-\vec x\,')\; ,
$$
for any choice of timelike coordinate $x^0$.

Note that in general, the (anti)commutator distributions can be
written as the difference of advanced and retarded
propagators~\cite{Lichnerowicz:1961}. 
In the Minkowski limit, the scalar
distribution $\Dp0(x,x';m^2)$ is the usual Pauli--Jordan commutator 
function.

For $s\geq1$, away from the gauge invariant
boundaries, the constraints~\eqn{const} and~\eqn{sticky} 
imply the (gamma-)traceless--transverse conditions
\be
\begin{array}{cc}
D.\phi=0\, ,&\\
&D.\psi=0=\g.\psi\, ,\\
D.\phi_\n=0=\phi_\r{}^\r\, ,&
\end{array}
\label{beauty}
\ee
which allow the field equations~\eqn{Proca}-\eqn{ES} to be rewritten
as
\be
\begin{array}{cc}
(\DB1-m^2)\,\phi_\m=0\, ,&\\
&(\frac{1}{2}\,\DF{3/2}+m)\,\psi_\m=0 \, ,\\
(\DB2-m^2+2\Lambda)\,\phi_{\m\n}=0\, .&
\end{array}
\label{captivate}
\ee
We must now write (anti)commutators for fields satisfying both~\eqn{beauty}
and~\eqn{captivate}. The latter requirement is just the analog of the
$s=0,1/2$ solutions given above. The former is easily imposed
using the (gamma-)\-traceless--transverse decompositions
\bea
\phi_\m^{\rm T}&=&\phi_\m-D_\m\,\frac{1}{D^2}\,D.\phi\, ,
\nn
\eea
\be
D.\phi^{\rm T}=0\, ;
\label{kane}
\ee
\bea
\psi_\m^{\TT}&=&\psi_\m
-\frac{1}{4}\,\g_\m\,\g.\psi
+D_{\wt\m}\,\frac{1}{3D^2+\L}\,(\sl D\g.\psi-4D.\psi)\, ,\nn
\eea
\be
D.\psi^{\TT}=0=\g.\psi^{\TT}\, ;
\label{T}
\ee
\bea
\phi^{\TT}_{\m\n}&=&
\phi_{\m\n}
-D_{(\m}\,\frac{2}{D^2+\L}\,(D.\phi_{\n)})^{\rm T}
-\,\frac{1}{4}\,g_{\m\n}\,\phi_\r{}^\r\nn\\&&
-\,D_{\{\m}D_{\n\}}\,
\frac{4}{D^2(3D^2+4\L)}\,\Big[D.D.\phi-\frac{1}{4}\,D^2\phi_\r{}^\r\Big]\, ,
\nn
\eea
\be
D.\phi_\m^{\TT}=0=\phi^{\TT}_{\;\r}{}^\r\, ;
\label{abel}
\ee
where a tilde over an index denotes its gamma-traceless part, {\it i.e.}
$X_{\wt \m}\equiv X_\m-\frac{1}{4}\,\g_\m \g.X$.
and $\{\cdot\cdot\}$ denotes the symmetric-traceless
part of any symmetric tensor, {\it i.e.} $X_{\{\m\n\}}\equiv
X_{(\m\n)}-\frac{1}{4}\,g_{\m\n}\,X_\r{}^\r$\, .

Therefore the (anti)commutators for spins $1\leq s\leq 2$ are given by
\bea
[\phi_\m(x),\phi_\n(x')]&\!\!=\!\!&i\,\Dp1_{\m\n}(x,x';m^2)
+\frac{i}{m^2}\,D_\m^x\, D_\n^{x'}\, \Dp0(x,x';m^2)\, ,
\label{onesies}\\\nn\\
\{\psi_\m(x),\ol\psi_\n(x')\}&\!\!=\!\!&
i\,\S3_{\m\n}(x,x';2m)-\frac{i}{4}\,\g_\m^x\,\S1(x,x';2m)\,\g_\n^{x'}\nn\\&&
\qquad+\;\frac{i}{3m^2+\L}\,\D_\m^x\,\S1(x,x';2m)\,\overleftarrow\D_\n^{x'}
\, ,
\\\nn\\
{}[\phi_{\m\n}(x),\phi_{\r\s}(x')]
&\!\!=\!\!&i\,\Dp2_{\m\n,\r\s}(x,x';m^2-2\L)
+\frac{2i}{m^2}\,D_\m^x\, D_\r^{x'}\, \Dp1_{\n\s}(x,x';m^2-2\L)\nn\\
&\!\!+\!\!\!&\frac{i}{m^2\,(3m^2-2\L)}\,\Big[2\,D^x_\m D_\n^x\,
D^{x'}_\r
D_\s^{x'}+m^2(\L-m^2)\,g_{\m\n}^x\,g_{\r\s}^{x'}\nn\\&&
+\,m^2\,D^x_\m D_\n^x\,g_{\r\s}^{x'}+m^2\,g_{\m\n}^x\,D^{x'}_\r
D_\s^{x'}
\Big]\;\Dp0(x,x';m^2-2\L)\, .\nn\\&&
\label{twosies}
\eea
For brevity, we have suppressed obvious symmetrizations $(\m\n)$ and
$(\r\s)$ on the right hand side of~\eqn{twosies}. 
The field equations~\eqn{captivate} have been used throughout to 
eliminate
any factors $D^2$ appearing in the (gamma-)traceless--transverse
decompositions~\eqn{kane}-\eqn{abel}, since the higher distributions are 
also onshell
\bea
(\DB{n}+m^2)\,\Dp{n}_{\m_1\ldots\m_n,\n_1\ldots\n_n}(x,x';m^2)&\equiv&0\, ,\\
(\DF{n/2}+m)\,\S{n}_{\m_1\ldots\m_n,\n_1\ldots\n_n}(x,x';m)\;\;&\equiv&0\, .
\eea 
The identity
$D_{\wt\m}^x\,\S1(x,x';2m)=\D^x_\m\,\S1(x,x';2m)$ is also useful. 
The
(anti)commutators~\eqn{onesies}-\eqn{twosies} are the
difference between advanced and retarded propagators~\cite{Lichnerowicz:1961}.
The higher distributions $\Dp1$, $\Dp2$ and $\S3$ above satisfy
\be
\begin{array}{cc}
D^\n_{x'}\,\Dp1_{\m\n}(x,x',m^2)=-D^x_\m\,\Dp0(x,x';m^2)\, ,&\\ \\
&\hspace{-2.2cm}\S3_{\m\n}(x,x';m)\,\overleftarrow\D^\n_{x'}=
-\D^x_\m\,\S1(x,x';m)\, ,\\
&\hspace{-2.2cm}\S3_{\m\n}(x,x';m)\,\g^\n_{x'}=
\g^x_\m\,\S1(x,x';m)\, ,\\ \\
D^\r_{x'}\,\Dp2_{\m\n,\r\s}(x,x',m^2)=-D^x_{(\m}\,\Dp1_{\n)\s}(x,x';m^2)\,
,&\\
g^{\r\s}_{x'}\,\Dp2_{\m\n,\r\s}(x,x',m^2)=
g_{\m\n}^x\,\Dp0(x,x',m^2)\, ,
\end{array}
\ee
with boundary conditions
$$
\frac{d}{dx'{}^0}\Dp1_{\m\n}(x,x')\Big|_{x^0=x'{}^0}=
\frac{g_{\m\n}}{\sqrt{-g}}
\,\delta^3(\vec x-\vec x\,')\;,\quad\
\Dp1_{\m\n}(x,x')\Big|_{x^0=x'{}^0}=0\, ,
$$
\be
\S3_{\m\n}(x,x')\Big|_{x^0=x'{}^0}=
\frac{g_{\m\n}\,\g^0}{\sqrt{-g}}
\,\delta^3(\vec x-\vec x\,')\; .
\label{ping}
\ee
$$
\frac{d}{dx'{}^0}\Dp2_{\m\n,\r\s}(x,x')\Big|_{x^0=x'{}^0}=
\frac{g_{(\m(\r}\,g_{\s)\n)}}{\sqrt{-g}}
\,\delta^3(\vec x-\vec x\,')\;,\quad\
\Dp2_{\m\n,\r\s}(x,x')\Big|_{x^0=x'{}^0}=0\, .
$$
Equipped with the above tools,
one can easily uncover the nonunitary regions. 
Starting with the
(anti)commutators~\eqn{onesies}-\eqn{twosies}, the aim is to determine
whether the distributions on their right hand sides have definite sign
in the dangerous lower helicity sectors.

For concreteness we work in the simple synchronous dS metric 
\be
ds^2=-dt^2+e^{2Mt}\,d\vec{x}^2\, , \qquad M\equiv \sqrt{\L/3}\ , \label{easy}
\ee
and concentrate on the equal time (anti)commutators of the
time components of the fields (and their time derivatives). While the
metric~\eqn{easy} does not cover the entire dS space, nor
is it real when continued to negative AdS values of $\L$,  these
disadvantages are outweighed by its simplicity (we will consider
more general coordinate systems later). 
Selecting the lowest helicity components by looking at
time components of fields,
a simple computation reveals that
\bea
[\phi_{0}(t,\vec x),\dot\phi_{0}(t,\vec{x}\,')]&=&
-\,\frac{\nabla^2}{m^2}\;\frac{i}{\sqrt{-g}}\,\delta^3(\vec x-\vec
x\,')\;,
\label{sump}\\
\nn\\
\{\psi_0(t,\vec x),\psi^\dagger_0(t,\vec x\,'\}&=&
-\,\frac{\nabla^2}{3m^2+\L}\;\frac{1}{\sqrt{-g}}\,\delta^3(\vec x-\vec x\,')\;,
\\\nn \\
{}[\phi_{00}(t,\vec x),\dot\phi_{00}(t,\vec{x}\,')]&=&
\frac{2\,\nabla^4}{m^2\,(3m^2-2\L)}\;\frac{i}{\sqrt{-g}}\,\delta^3(\vec
x-\vec x\,')\;, \label{sung}
\eea
where $\nabla^2\equiv g^{ij}\d_i\d_j =e^{-2Mt}\,\vec \d\,^2$ is a negative
operator. The final equation~\eqn{sung} agrees with the
detailed massive $s=2$ Dirac analysis presented
in~\cite{Higuchi:1987py}. Our derivation only requires writing out the Laplace
and Dirac operators explicitly in the metric~\eqn{easy} and using the field 
equations~\eqn{captivate} 
to maximally eliminate time derivatives, after which the
boundary conditions~\eqn{ing} and~\eqn{ping} may be applied. 

The interpretation of equations~\eqn{sump}-\eqn{sung} is
as follows: Positivity of the distributions on the right hand sides
is completely determined by the respective denominators $m^2$, $3m^2+\L$ and
$3m^2-2\L$, precisely the factors appearing in the
Bianchi identities of the previous Section. For the various spins, we
learn:
\begin{itemize}
\item \underline{Spin~1:\ } The model is unitary in the entire
$(m^2,\L)$ half-plane. A line of gauge invariant models emerges at
$m^2=0$ (the same value as in flat space).
\item \underline{Spin~3/2:\ } The model is unitary in the region
$m^2>-\L/3$ which includes the Minkowski background. 
Strictly massless, gauge invariant unitary models
are found along the AdS line $m^2=-\L/3$. The region $m^2<-\L/3$ is
non-unitary. In contrast to flat space, the $m^2=0$ theories are {\it massive}
when $\L\neq0$ and even non-unitary for negative (AdS) values of 
$\L$.  
\item \underline{Spin~2:\ } Models with $m^2>2\L/3$ are unitary.
There are now two lines of gauge invariant theories; the usual
linearized cosmological Einstein theory at $m^2=0$ and a partially massless 
theory~\cite{Deser:1983tm} at $m^2=2\L/3$. Both are unitary
but the region $m^2<2\L/3$ is not.
\end{itemize}
These conclusions are depicted in Figure~\ref{regions}. 

Finally, as promised, we address the concern that, strictly, the
metric~\eqn{easy} applies only to dS.
On the one hand, given that (i) the final results are a function
of the real variable $\L$ only and (ii) the picture presented
here is backed up by the emergence of Bianchi identities, there can be
no doubt of its correctness. However, for complete certainty, we
repeat, as an example, the $s=3/2$ computation in the metric
($M\equiv\sqrt{\L/3}$)
\be
ds^2=-dt^2+\cosh^2(Mt)\,\Big\{dr^2+\frac{1}{M^2}\,\sin^2(Mr)\,(d\theta^2+
\sin^2\theta\,d\phi^2)\Big\}\, . \label{buoy}
\ee
Upon rescaling $r\rightarrow\rho/M$, the three-metric 
$d\Omega^2=d\r^2+\sin^2\r\,(d\theta^2+\sin^2\theta\,d\phi^2)$ 
is seen to describe a unit
three-sphere. We prefer the initial parametrization, however, since for pure
imaginary values  of $M$, the cosmological constant $\L$ is  negative
and the metric continues to AdS.
Performing a calculation analogous to the one above we find 
\bea
\{\psi_0(t,r,\theta,\phi),\psi^\dagger_0(t,r',\theta',\phi')\}&=&
\nn\\&&\hspace{-4.2cm} 
\frac{\cosh^{-2}(Mt)\,(-\,^{(3)}\!D^2-\L/4)}{3m^2+\L}
\;\frac{1}{\sqrt{-g}}\,\delta(r-r')
\,\delta(\theta-\theta')\,\delta(\phi-\phi')\;.\nn\\
\eea
The operator 
${}^{(3)}\!D^2$ is the square of the intrinsic 3-dimensional covariant
derivative (Laplace--Beltrami operator) 
acting on a spinor. In dS the operator
$-\,^{(3)}\!D^2-\L/4$ is not manifestly positive. 
However (in our parametrization) the eigenvalues
of ${}^{(3)}\!D^2$ acting on spinors are $(\L/3)(-l(l+2)+1/2)$  
with $l\geq 1/2$ (see,
{\it e.g.}~\cite{Higuchi:1987py}),  
and the highest eigenvalue is precisely $-\L/4$. Hence the operator
$-\,^{(3)}\!D^2-\L/4$ is indeed positive and in dS we may draw
precisely the conclusions given above. Now, continuing the
metric~\eqn{buoy} to AdS space the same result holds for the local
anticommutator except the 3-space is a hyperboloid. Nonetheless,
(assuming we can neglect spatial boundary terms), both
$-{}^{(3)}\!D^2$ and $-\L/4$ are now separately positive, and unitarity is
determined by the sign of the denominator $3\L+m^2$. This concludes
our derivation of the unitarily forbidden regions for spins $s\leq2$.

We emphasize that once one knows the gauge lines and their
corresponding Bianchi identities, our unitarity results in fact follow
by inspection: Whenever a coefficient in a massive constraint vanishes
and then becomes negative, all the corresponding lower helicity modes are first
excised by the accompanying gauge invariance and thereafter reemerge
with opposite norms. Therefore, starting from the unitary Minkowski
region, it is easy to map out the unitarily allowed and forbidden
regions, as shown in Figure~\ref{regions}. Furthermore, for higher spin
partially massless theories to be unitary, the ordering criterion for
the gauge lines, discussed in the introduction, must hold. A simple
example is provided by the $s=2$ strictly massless ($m^2=0$, 
linearized graviton) 
theory for $\L>0$:  To reach it
starting from the unitary Minkowski region, one must pass through the
unitarily forbidden region $0<m^2<2\L/3$.
Nonetheless, the theory is unitary, since the highest helicities
$\pm2$ are
left untouched by the unitarity flip of the helicity $0$ mode across
the $m^2=2\L/3$ gauge line. 

\section{\hspace{-.3cm}Partially$\!$ Massless$\!$ 
Spin$\!$~2: Canonical$\!$ Analysis}

\label{whale}

At the partially massless dS
boundary, $m^2=2\L/3=2M^2$, we showed that the 
scalar constraint~\eqn{sticky} is a Bianchi identity; as such 
it removes 2
DoF, leaving 4 physical DoF corresponding to helicities $\pm2,\pm1$. 
In this Section we prove this claim via an explicit canonical
analysis\footnote{A detailed canonical analysis of massive $s=2$ for
general $m^2$ is given in~\cite{Deser:2002}; an early attempt can be found 
in~\cite{Bengtsson:1995vn}.}. Our method is similar to that originally
used to prove the stability of massless
cosmological gravity~\cite{Abbott:1982ff}.

A possible starting point is the second order massive $s=2$ 
action in equation~\eqn{ramit}. Equivalently (and much simpler) one can
begin with the first order ADM form of cosmological Einstein gravity,
\bea
S_{\L+E}&=&\int d^4x\;
\Big[\;\pi^{ij}\,\dot g_{ij}
+N\,\sqrt{g}\;\Big({}^{(3)}\!R-2\L\Big)+2N_iD_j\pi^{ij}
\nn\\&&\qquad\qquad\quad
+\frac{N}{\sqrt{g}}\,\pi^{ij}\pi^{lm}\,
\Big(\frac{1}{2}\,g_{ij}g_{lm}-g_{il}g_{jm}\Big)
\Big]\, ,\label{einst}
\eea
then linearize around a dS background and add by hand 
an explicit mass term. Here $g$ is the determinant of the
3-metric $g_{ij}$ and $N\equiv(-g^{00})^{-1/2}$, $N_{0i}\equiv
g_{0i}$.
We take the synchronous dS metric~\eqn{easy},  
denoted $ds^2=-dt^2+\ol g_{ij}dx^i dx^j$ in this Section,
reserving $g_{ij}$ for the dynamical 3-metric which we linearize as
\be
g_{ij}\equiv\ol g_{ij}+\phi_{ij}\, ,\qquad\ol g_{ij}\equiv f^2(t)\,\delta_{ij}
\; ,\qquad f(t)\equiv e^{Mt}\, .
\ee
The remaining fields are linearized as
\be
\pi^{ij}\equiv\ol\pi^{ij}+P^{ij}\;,\quad\ol
\pi^{ij}\equiv-2Mf\,\delta^{ij}\;,\qquad
N\equiv1+\wt n\, .
\ee
(The background metric is block diagonal so no expansion is needed for
$N_i$.) In terms of these deviations, the mass term is
\be
S_m=-\frac{m^2}{4}\,\int d^4x\,\sqrt{\ol g}\,
\Big(\phi_{\m\n}\phi_{\r\s}\ol g^{\m\r} \ol g^{\n\s}-(\phi_{\m\n}\ol
g^{\m\n})^2\Big)\, ,\label{mass}
\ee
here $\phi_{0i}\equiv N_i$, $\ol g\equiv\det \ol g_{ij}=-\det\ol
g_{\m\n}$ and $\phi_{00}\equiv g_{00}+\ol g_{00}=-(1+N^2)$.
The final action is the sum $S=S_{\L+E}+S_m$, discarding any
terms of higher than quadratic order in (dynamical) fields.

Notice that the only explicit time dependence of the integrand of~\eqn{mass} is
through $f^{-1}$. Indeed it proves useful to make the field redefinition
\bea
\phi_{ij}\equiv f^{1/2}\,h_{ij}\; ,\qquad
P^{ij}\equiv f^{-1/2}\,p^{ij}\, .
\nn
\eea
\be
N_i\equiv  f^{1/2}\,n_i\;,\qquad \wt n=f^{-3/2}\,n\, .
\label{redef}
\ee
The cost is an extra contribution generated by the symplectic
term of~\eqn{einst}, $P^{ij}\dot\phi_{ij}$ $\rightarrow$ 
$p^{ij}\dot h_{ij} + (M/2) p^{ij}h_{ij}$. A dividend is that the only
explicit time dependence in what follows will be through $\nabla^2\equiv
\ol g^{ij}\d_i\d_j\equiv f^{-2}\nabla_0^2$. Index contractions are 
just with $\delta_{ij}$, all quantities are now in $(3+1)$ form.

Next examine the mass term
\be
S_m=\int d^4x\,\Big[
-\frac{m^2}{4}\,(h_{ij}^2-h_{ii}^2)+\frac{m^2}{2}\,N_i^2+m^2
\,n\,h_{ii}\Big]\ .
\ee
Were it not for the term proportional to $N_i^2$, the field $N_i$ would 
be a  Lagrange multiplier for 3 constraints (as is the case for
the $m=0$ strictly massless theory). 
Instead, when $m\neq0$ we must   integrate out $N_i$
via its algebraic equation of motion. The field $n$, however, only
appears linearly and remains a Lagrange multiplier for the constraint
\be
\Big[\sqrt g\,({}^{(3)}\!R-2\L)+m^2\,h_{ii}
+\frac{1}{\sqrt{g}}\,\pi^{ij}\pi^{lm}\,
(\frac{1}{2}\,g_{ij}g_{lm}-g_{il}g_{jm})\Big]_{\rm \sss linearized}=0\, .
\label{connie}
\ee
For generic values of $(m^2,\Lambda)$, this constraint eliminates 
one degree of freedom from the 6 pairs $(p^{ij},h_{ij})$ 
leaving 5 physical helicities $(\pm2,\pm1,0)$.
Our aim now is to show that 
a (single) further constraint emerges at the gauge invariant value
$m^2=2M^2$.

We decompose the
fields $h_{ij}$ and $p_{ij}$ according to their helicity
and drop (for now) the helicity 
$\pm2$ traceless-transverse $(h_{ij}^{tt},p_{ij}^{tt})$ {\it and}
helicity $\pm1$ transverse $(h_{i}^{t},p_{i}^{t})$ modes
since they manifestly decouple from the 2 remaining helicity 0 modes
(to the quadratic order used here). The latter are defined
by the projection
\be
h_{ij}=\frac{1}{2}\,\Big(\delta_{ij}-\frac{\d_i\d_j}{\nabla_0^2}\Big)\,h_T
+\frac{\d_i\d_j}{\nabla_0^2}\,h_L\, ,
\ee
and similarly for $p_{ij}$. (Note that under the integral 
$\int A_{ij}\,B_{ij}=
\frac{1}{2}\,\int A_T B_T+\int A_L\,B_L$.)
In terms of these variables the linearized
constraint is
\be
{\cal
C}\equiv\nabla^2h_T+2M\,(p_T+p_L)-(m^2-2M^2)\,(h_T+h_L)=0\,
.\label{twang} 
\ee
Note that the leading term comes from the linearized 3-dimensional
curvature scalar and that both the cosmological $-2\L\sqrt{g}$ and
momentum squared terms in~\eqn{connie} contribute to the
final term in~\eqn{twang}, which vanishes on the critical gauge
line $m^2=2M^2$. Henceforth we concentrate on the critical case 
and eliminate $m^2$ via this 
relation.

Next we write out the quadratic action $S=S_{E+\L}+S_m$ 
remembering the constraint~\eqn{twang} which we solve as
\be
p_T=-p_L-\frac{\nabla^2}{2M}\;h_T\, .
\ee
Observe that the symplectic terms become
\be
p_{ij}\,\dot h_{ij}=\frac{1}{2}\,p_T\dot h_T+ p_L\dot
h_L= p_L\dot q- \frac{1}{4}\,h_T\,\nabla^2\,h_T\; ,\qquad
q\equiv h_L-\frac{1}{2}\,h_T\, .
\ee
(suppressing integrations throughout).
Upon eliminating $h_L=q+h_T/2$ in favor of $q$ and $h_T$, the action
depends only on the 3 variables $(p_L, q,h_T)$ and its most general
form is
\be
S(p_L, q,h_T)-  p_L\,\dot q= \frac{1}{2}\,A\, h_T^2\,+\,B\,h_T\,+\,C\ ,
\ee
where $A$ is constant, $B$ linear and $C$ quadratic in $(p_L,q)$. 
If $A=0$, we have
an additional constraint $B(p_L,q)=0$ and no zero helicity PDoF
remain, whereas for non-zero $A$, 
$h_L$ can be removed via its algebraic
field equation leaving behind one zero helicity DoF.
In fact, $A$ does vanish on the
critical line $m^2=2\L/3$ so this model 
describes helicities $(\pm2,\pm1)$ only. 
Indeed, a lengthy  calculation yields
\be
S(p_L, q,h_T)-  p_L\,\dot q= 
\frac{1}{2}\,\Big(\frac{p_L}{M}-q\Big)\nabla^2\,h_T
\,+\,\Big(\frac{p_L}{M}-q\Big)
\Big[\!\Big(\frac{\nabla^2}{m^2}-\frac{3}{2}\Big)\,p_L
-\Big(\!\nabla^2-m^2\Big)\,q\Big]\ .
\ee 
As claimed, $A=0$ and even the zero helicity Hamiltonian vanishes once
the Lagrange multiplier $h_T$ is integrated out. 

Let us now examine the remaining helicities $(\pm2,\pm1)$. A series of
canonical transformations yields a simple action
\be
S_{(\pm2,\pm1)}=\sum_{\ve=(\pm2,\pm1)}\,\Big\{ p_{\ve}\,\dot q_\ve
-
\frac{1}{2}\,
\Big[ \,p_\ve^2\ + \
q_\ve\,\Big(\!-\nabla^2-\frac{M^2}{4}\Big)\,q_\ve \
\Big]\Big\}\, .
\label{hell}
\ee
Notice again, all time dependence is through $\nabla^2$ in the Hamiltonian.
The field equations are
\be
p_\ve=\dot q_\ve\ , \qquad
\Big(-\frac{d^2}{dt^2}+\nabla^2-\frac{M^2}{4}\ \Big)\,q_\ve=0\ .
\label{picky}
\ee
The covariant field equation~\eqn{captivate} evaluated at
$m^2=2M^2$ 
is $(D^2-4M^2)\,\phi_{\m\n}=0$. Consider, for example, helicities
$\pm 2$, for which $\d^i\phi_{ij}=0=\phi_i{}^i$. In this frame
the transverse-traceless part of the covariant field equation reads
\be
\Big(-\frac{d^2}{dt^2}\ + \ M\,\frac{d}{dt}\ +\
\nabla^2\ \Big)\,\phi_{ij}=0\, .\label{icky}
\ee
The action~\eqn{hell} was obtained by the same rescaling as
in~\eqn{redef},
namely, $\phi_{ij}=f^{1/2}\,q_{ij}$. The factor $f^{1/2}$ is 
precisely the integrating factor which removes the single time
derivative from equation~\eqn{icky} at the cost of a term $-M^2/4$,
{\it i.e.} equations~\eqn{icky} and~\eqn{picky} are identical
(helicities $\pm1$ agree via a similar calculation).

Stability of the partially massless theory
requires that it possess a conserved, positive, energy function. 
The latter can be
obtained by an argument similar to that  given 
in~\cite{Abbott:1982ff} for the
strictly
massless $s=2$ theory:
The Hamiltonian in~\eqn{hell} is not conserved because of the explicit
time dependence of $\nabla^2$. However, inside the intrinsic dS 
horizon at $(fMx^i)^2=1$, the background metric~\eqn{easy}  
possesses a timelike Killing vector
\be
\xi^\m=(-1,Mx^i) \Longrightarrow \xi^2=-1+(fMx^i)^2\ .
\ee
Therefore, the energy associated
with time evolution in this Killing direction
\be
E=T^0{}_\m\xi^\m=
H-Mx^i\,\Big[p_\ve\,\d_i q_\ve-\frac{1}{2}\,\d_i\,(p_\ve q_\ve)\Big]
\ee
satisfies $\dot E=0$ ($H$ is the Hamiltonian in~\eqn{hell} and
we have suppressed the sum over helicities $\ve$). Furthermore,
writing out $E$ explicitly and relabeling the variable
$p_\ve\rightarrow p_\ve+(3M/2)q_\ve$ gives
\be
E=\frac{1}{2}\,\Big(\wh x^i\,p_\ve\Big)^2
+\frac{1}{2}\,\Big(f^{-1}\d_i\,q_\ve\Big)^2 -f M |x|\,\Big(\wh
x^i\,p_\ve\Big)\,\Big(f^{-1}\d_i\,q_\ve\Big)
+\frac{1}{2}\,(2M^2)\,q_\ve^2\, ,
\ee
with $x^i\equiv |x|\,\wh x^i$.
The last (mass) term is manifestly positive and the first three terms
are positive by the triangle equality whenever
\be
f\,M \,|x|<1\ ,
\ee
that is, inside the physically accessible region.

A final interesting feature of the partially massless $s=2$ theory is
null propagation. 
The dS metric is conformally flat and it can be
shown that the $m^2=2\L/3$ theory propagates on its null 
cone~\cite{Deser:1983tm}.
An interesting open question is whether any of the $s>2$
partially massless theories which we discuss next,  
share this behavior.

\section{Higher Spins} 

\label{dolphin}

Having seen that the $s=2$ field equation ${\cal G}_{\m\n}$ yields
both single divergence, $D.{\cal G}_{\n}$, and double divergence, 
$D.D.{\cal G}$, Bianchi identities we are led to inquire whether even
higher divergence Bianchi identities occur for $s>2$. The answer
is yes. This implies that, in addition to
the usual massive and strictly massless possibilities, 
a spin $s$ field in (A)dS
can be partially massless with propagating helicities 
$(\pm s,\pm(s-1),\ldots,\pm(s-t))$ ($t<s$).
We first demonstrate this claim by simple counting arguments and then
write explicit gauge invariances and Bianchi identities for  
$s=5/2,3$ examples.

\subsection{Higher Spin Bosons}

Define 
the number of components $\sy s$ of a symmetric $s$-index tensor 
$\phi_{\m_1\ldots\m_s}$ 
\be
\sy s\equiv
\left\{
\begin{array}{cl}
\frac{(s+1)(s+2)(s+3)}{3!}&s>0\\ \\
0&\mbox{otherwise}\, .
\end{array}
\right.
\ee 
Then a traceless symmetric tensor has $\syt s$ components
\be
\syt s=\sy s-\sy{s-2}\; (=(s+1)^2 \ , s\geq0)\ , 
\ee 
while a doubly traceless ($0=\phi^\r{}_\r{}^\s{}_{\s\m\ldots}$)
one has
\be
\sytt s=\sy s- \sy{s-4}\;  (=2(s^2+1)\ ,s\geq1).
\ee

\subsubsection*{Strictly Massless Bosons}

First consider massless bosons.
The field content is a doubly traceless $s$ index symmetric tensor
enjoying gauge invariances with an $s-1$ index traceless gauge
parameter: 
\begin{center}
\begin{tabular}{|rl|c|}
\hline
&Fields&$\ss \sytt s$ \\ \hline \hline 
$\!-\hspace{-.3cm}$ &Gauge&$\ss 2.\syt{s-1}$\\
\hline
\end{tabular}
\end{center}
Since $2\syt{s-1}$ DoF can be gauged away, the PDoF for $s\geq1$ are  
$\sytt s-2\syt{s-1}=2$, corresponding to
helicities $\pm s$.

\subsubsection*{Massive Bosons}

For  massive $s>2$ theories, the massless field content must be augmented 
by a set of traceless symmetric auxiliary fields, 
$(\chi,\ldots,\chi_{\m_1\ldots\m_{s-3}})$.
Each divergence of the 
$s$-index symmetric field equations is a constraint whenever the remaining
open indices are traceless:
\begin{center}
\begin{tabular}{|rl|cl|}
\hline
&Fields&$\ss \sytt s$&\\ 
$\!+\hspace{-.3cm}$ &Auxiliaries&$\ss \syt 0+\cdots+\syt {s-3}$&\\ 
\hline\hline
$\!-\hspace{-.3cm}$ &Constraints&$\ss\syt 0+\cdots+\syt {s-3}$&
$\hspace{-.37cm}\ss+\,\syt{s-2}+\syt{s-1}$
\\\hline
\end{tabular}
\end{center}
and the PDoF are $\sytt s-\syt{s-2}-\syt{s-1}=2s+1$, the sum
of all helicities.

\subsubsection*{Partially Massless Bosons} 

For partially massless higher spin theories the field content is the
same as the massive case but new Bianchi identities appear.
There are as many of these as
possible divergences of
the $s$-index symmetric field equations.
On a gauge line where a constraint
with $t$ divergences becomes a Bianchi identity we have:
\begin{center}
\begin{tabular}{|rl|cl|}
\hline
&Fields&$\ss \sytt s$&\\ 
$\!+\hspace{-.3cm}$ & Auxiliaries&
$\ss \syt 0+\cdots+\syt{s-t}+\syt{s-t+1}
+\cdots+\syt {s-3}$
&\\
\hline\hline
$\!-\hspace{-.3cm}$ & Constraints&
$\ss \phantom{\syt 0 +\cdots +2.\syt{s-t}+}
\!\!\!\!\syt{s-t+1}
+\cdots+\syt {s-3}$ &
$\!\!\!\!\!\!\ss +\syt{s-2}+\syt{s-1}$\\
$\!-\hspace{-.3cm}$ & Gauge&
$ 
\hspace{-1.8cm}\ss2.\syt{s-t}$&\\\hline
\end{tabular}
\end{center}
The $t$-divergence Bianchi identity replaces the set of constraints with 
$t,t+1,\ldots,s$ divergences and the corresponding gauge invariance
removes $2.\syt{s-t}$ DoF.
Therefore we find
$
2t
$
PDoF along with a set $(\syt 0 +\cdots + \syt {s-2-t})$ of leftover
auxiliary fields.
When the new Bianchi identity is the
scalar one with the maximal number of 
divergences $t=s$, there are $2s$ PDoF and 
only the massive helicity $0$ mode is removed. For the vector Bianchi with
$t=s-1$ there are $2s-2$ PDoF because helicities $(0,\pm1)$ are excised.
For even lower values of $t$, the sum over leftover auxiliaries
$\syt 0 +\cdots + \syt {s-2-t}$
is non-empty:
In the strictly massless case $t=1$, there $2$ PDoF (helicities $\pm
s$) and all the auxiliaries $\syt 0 +\cdots + \syt {s-3}$ remain.
However, at least in this case, it is obvious that they decouple.
Less obvious is whether they decouple for the
partially massless $s\geq4$ theories. If not, we must
then ask how many PDoF they represent and whether they are unitary
excitations, questions clearly inaccessible via the simple counting
arguments presented here.

\subsection{Higher Spin Fermions}

For brevity define $\s\equiv s-1/2$.
The (real) components of a $\s$-index, symmetric, gamma-traceless 
(Majorana) field are denoted 
\be
4.\syg\s=4\,[\sy\s-\sy{\s-1}]\;\;(=2(\s+1)(\s+2)\ , \s\geq0) \ ,
\ee 
and a
traceless gamma-tracelessness one ($\g.\psi^\r{}_{\r\m_2\ldots\m_\s}=0$) has
\be
4.\sygt\s=4\,[\sygt\s-\sygt{\s-3}]\;\;(=2(3s^2+3s+2)\ , \s\geq0) 
\ee 
components. 

\subsubsection*{Strictly Massless Fermions}

For massless half integer $s=\s+1/2$ the 
field content is a traceless gamma-traceless, symmetric $\s$-index
spinor. The correct number of PDoF is ensured by gauge invariances with
a gamma-traceless, symmetric $(\s-1)$-index spinor parameter:
\begin{center}
\begin{tabular}{|rl|c|}
\hline
 $1/2$& Fields&$\ss 4.\sygt \s$ \\ \hline 
$-1/2$& Gauge&$\ss 4.\syg{\s-1}$\\
\hline\hline
$-\;$& Gauge&$\ss 4.\syg{\s-1}$\\
\hline
\end{tabular}
\end{center}
To explain: we begin with $4.\sygt \s$ fields, of which gauge
invariance removes $4.\syg{\s-1}$ components. However, before
excising further DoF via residual invariances we must impose the
projector field equations (recall that the Dirac equation
divides the DoF of a spinor by 2). Thereafter residual gauge invariances
remove $4.\syg{\s-1}$ further DoF. Hence the PDoF count is 
$2\,[\sygt\s-3\syg{\s-1}]=2$, corresponding to helicities $\pm s$.

\subsubsection*{Massive Fermions}

In addition to the same fields as for massless fermions, 
there are traceless auxiliary spinor fields
$(\chi,\ldots,\chi_{\m_1\ldots\m_{\s-2}})$ when $s\geq3/2$. 
All possible divergences on $\s$
vector indices of the field equations yield constraints when 
the leftover indices are gamma-traceless:
\begin{center}
\begin{tabular}{|rl|c|}
\hline
 $1/2$& Fields&$\ss 4.\sygt \s$ \\ \hline 
 $1/2$& Auxiliaries&$\ss 4.\syt{0}+\cdots+4.\syt{\s-2}$\\
\hline\hline
$-\;$& Constraints&$\ss 4.\syg{0}+\cdots+4.\syg{\s-1}$\\
\hline
\end{tabular}
\end{center}
Adding these up yields $2\s+2=2s+1$ PDoF and all helicities
$(\pm s,\ldots,\pm1/2)$ propagate.

\subsubsection*{Partially Massless Fermions}

The field content is identical to massive case but now suppose that 
$t$ divergences of the field equations yield a Bianchi identity. We
have:
\begin{center}
\begin{tabular}{|rl|cl|}
\hline
 $1/2$& Fields&$\ss 4.\sygt \s$ &\\ \hline 
 $1/2$& Auxiliaries&$\ss 4.\syt{0}+\cdots+4.\syt{\s-2}$&\\ 
$-1/2$& Gauge&$\ss 4.\syg{\s-t}$&\\
\hline\hline
$-\;$& Constraints&&$\ss \hspace{-1cm}4.\syg{\s-t+1}+\cdots+4.\syg{\s-1}$\\
\hline
$-\;$& Gauge&$\ss 4.\syg{\s-t}$&\\
\hline
\end{tabular}
\end{center}
The sum is $2t+2\,[\syt0+\cdots+\syt{\s-t-1}]$. Again we find $2t$
propagating helicity states $(\pm s,\ldots,\pm(s-t))$
along with leftover auxiliaries which decouple on the strictly
massless $(t=1)$ gauge line. (Just as in the bosonic case, 
for partially massless spins
$s\geq7/2$, little is known about the role of these auxiliaries). 

We next present the explicit $s=5/2,3$ examples. For $s=5/2$
we will exhibit one new gauge line resulting from the fermionic 
generalization of the $s=2$ double divergence Bianchi
identity~\eqn{sticky}. The resulting helicity $(\pm5/2,\pm3/2)$ 
partially massless theory
lives in an AdS region where helicities $\pm3/2$ have negative norms
and therefore fails to be unitary. In contrast, for $s=3$ we find
both double and triple divergence Bianchi identities corresponding to
unitary, partially massless, dS theories of helicities $(\pm3,\pm2)$ and
$(\pm3,\pm2,\pm1)$.

\subsection{Spin~5/2}

The $s=5/2$ spinorial field equation has two open indices, so as for
$s=2$, there are two possible Bianchi identities; they appear along
the AdS gauge lines $m^2=-4\L/3$ and $m^2=-\L/3$ (see
Figure~\ref{regions}). 
The former is the
strictly massless theory with helicities $\pm5/2$ whereas the novel gauge
invariance of the latter removes only the lowest $\pm1/2$ leaving
helicities $(\pm5/2,\pm3/2)$. Since the massless 
gauge lines all 
lie in AdS (just as for their $s=3/2$ counterpart), the $(m^2,\L)$ half-plane
is divided into 3 regions. Only the $m^2>-4\L/3$ one including
Minkowski space, is unitary.

The $s=5/2$ action and field equations are 
\be
{\cal L}=-\,\sqrt{-g}\;\ol\psi^{\m\n}\,{\cal R}_{\m\n}
-\,\sqrt{-g}\;\ol\chi\,{\cal R}_5\, ,
\label{rosella}
\ee
\bea
{\cal R}_{\m\n}&=&(\DF{5/2}-2\Sl D)\,\psi_{\m\n}+g_{\m\n}\,\g.D.\psi
+(D_{(\m}\g_{\n)}-\frac{1}{2}\,g_{\m\n}\,\Sl D)\,\psi_\r{}^\r\\
&+&m\,(\psi_{\m\n}-2\,\g_{(\m}\g.\psi_{\n)}-\frac{1}{2}\,g_{\m\n}\,
\psi_\r{}^\r)
-\frac{5}{12}\,\mu\,g_{\m\n}\,\chi\, ,
\label{RR}\\
{\cal R}_5&=&-\a\,(\Sl D-3m)\,\chi-\frac{5}{12}\,\mu\,\psi_\r{}^\r\, .
\label{R5}
\eea
They can be derived by minimally coupling the flat space equations of
Appendix~\ref{5/2fl} to the cosmological background. The former were
obtained by KK descent from their massless counterparts in $d=5$.
Minimal coupling alone does not provide equations of motion
describing the $6=2s+1$ massive PDoF: An additional non-minimal
coupling contained by the term $\mu \,\ol \chi\, \psi_\r{}^\r$ 
is necessary. In fact, to achieve a proper set of constraints requires
fixing the auxiliary coupling to
\be
\mu^2=\frac{12\a}{5}\,(m^2+4\L/3)\ .
\label{mu}
\ee
Here we encounter a new subtlety. Implicitly we have so far assumed
that physical models live in the half plane $m^2>0$, since for
fermions negative $m^2$ implies a non-hermitean mass term, and
for bosons, one that is unbounded below. While there are regions
with both $m^2<0$ and the correct sign for anticommmutators, the dynamics 
is non-unitary there. Therefore we continue to require
$m^2>0$ and
examine the relation~\eqn{mu}. Since hermiticity of the action~\eqn{rosella}
demands $\mu^2>0$, we find two regions: (i) $m^2>-4\L/3>0$,
(ii) $0<m^2<-4\L/3$. Up until now, $\a$ was a free parameter which we
could set to $\pm1$.
In the region (i), $m^2+4\L/3>0$ so we must take $\a=+1$. In region
(ii), hermiticity of the action can be maintained at the cost of
changing the sign of $\a$ to  $\a=-1$ (the actions are then different
in each region).  In either case, we will find that region
(ii) is unitarily forbidden.

Before continuing, it is interesting to
compare these difficulties to $s=3/2$ and the problem of constructing
dS supergravities~\cite{Pilch:1985aw}. As we have shown, $s=3/2$ is
unitary 
for $m^2\geq -\L/3\geq0$ and the boundary is the strictly massless
AdS theory corresponding to cosmological supergravity. 
As one follows the massive theory into dS, the canonical
anticommutators  all have the correct sign for unitary representations.
In fact, keeping $\L>0$, there is no obstruction at the level of
anticommutators to continuing to $m^2<0$ until 
the branch of the line $m^2=-\L/3$ with $m^2<0<\L$ is reached.
The theory there is formally
supersymmetric ({\it i.e.} strictly massless) 
but the action is no longer hermitean, 
which is a example of the general statement that dS
supergravities do not exist~\cite{Pilch:1985aw}. One might speculate that this
clash is generic to higher spin fermions.
 
Returning to the $s=5/2$ field equations, for generic $(m^2,\L)$
the constraints
\bea
{\cal C}_{\wt \n}&\equiv&
D.{\cal R}_{\wt \n}+\frac{1}{4}\,m\,\g.{\cal R}_{\wt \n}
\label{C51}
\\
&=& -\frac{5}{4}\,(m^2+4\L/3)\,\g.\psi_{\wt \n}
-\frac{5}{12}\,\mu\,D_{\wt \n}\,\chi\, ,\label{coleslaw}\\
{\cal C}&\equiv&
D.{\cal C}+
\frac{5}{16}\,(m^2+4\L/3)\,{\cal R}_\r{}^\r
-\frac{5}{16}\,\a\mu\,(\Sl D+3m)\,{\cal R}_5
\label{c1}\\&=&
-\frac{10}{3}\,\mu\,(m^2+\L/3)\,\chi
\ ,\label{C52}
\eea
ensure that the model describes $6=2s+1$ massive PDoF. 
Along the AdS lines
\be
m^2=-4\L/3\, \qquad m^2=-\L/3
\ee
the constraints~\eqn{coleslaw} and~\eqn{C52} transmute to
Bianchi identities with respective (distinct) gauge invariances
\bea
\delta\psi_{\m\n}=D_{(\m}\varepsilon_{\wt
\n)}+\frac{1}{2}\,\sqrt{\frac{-\!\L\,}{3}} \,\g_{(\m}\,\varepsilon_{\wt
\n)}\, ,&&\delta\chi=0\, ,
\label{ar}\\
\delta\psi_{\m\n}=D_{(\m}D_{\wt \n)}\,\varepsilon
+\frac{5\L}{16}\,g_{\m\n}\,\varepsilon\, ,\qquad&&
\delta\chi=-\frac{1}{8\a}\,\sqrt{15\a\L}\,(\Sl D+\sqrt{-3\L})
\,\varepsilon \, .\qquad\label{ra}
\eea
The vector-spinor Bianchi identity~\eqn{C51} at $m^2=-4\L/3$
implies strict masslessness (propagating helicities $\pm5/2$) 
since its invariance removes helicities
$(\pm3/2$, $\pm1/2)$. Notice also that $\mu=0$ on the strictly massless
line so, as claimed above, the spinor auxiliary $\chi$ decouples there.
The novel spinor Bianchi identity~\eqn{C52} at $m^2=-\L/3$
and invariance~\eqn{ra}
removes helicities $\pm1/2$ leaving
a partially massless theory of helicities $(\pm5/2,\pm3/2)$.

Once again, the coefficients $(m^2+4\L/3)$
and $(m^2+\L/3)$ appearing in the constraints~\eqn{C51} and~\eqn{C52} 
control the positivity of equal time anticommutators.
Therefore, since the 
gauge lines all 
lie in AdS, the $(m^2,\L)$-plane
is divided into 3 regions; only the one including Minkowski space
$m^2>-4\L/3$ is unitary.
Although the strictly massless, AdS, $m^2=-4\L/3$ theory is unitary, the
partially massless one is not, as
it fails the line ordering requirement: Starting from the unitary Minkowski
region where all norms are positive,
one would like first to traverse the line $m^2=-\L/3$, but that is only
possible in dS with negative $m^2$ (imaginary values of $m$ violate
hermiticity of the action and unitary evolution). Crossing the AdS
strictly massless line $m^2=-4\L/3$ first flips the norm of both lower
helicities
$(\pm3/2,\pm 1/2)$ so the partially massless AdS theory cannot be
unitary. Ironically, were negative values of $m^2$ not
prohibited, we could traverse the lines in the correct order in
dS. This observation lends weight to the speculation that the unitarity
difficulties of partially massless theories are peculiar to half
integer spins. Indeed, in the next Section we exhibit the bosonic
$s=3$ example, which enjoys two partially massless unitary  dS lines.

\subsection{Spin~3}

Spin~3 is the first example of a system 
with two new Bianchi identities over and above the usual one at
$m^2=0$. 
The action and field equations (with flat space limits derived by KK
descent in Appendix~\ref{3fl}) are
\be
{\cal L}=\,\frac{1}{2}\,\sqrt{-g}\,\phi^{\m\n\r}\,{\cal G}_{\m\n\r}
\,-\,\frac{3}{8}\,\sqrt{-g}\,\chi\,{\cal G}_{5}\ ,
\ee
\bea
{\cal
G}_{\m\n\r}&=&(\DB3-m^2+16\L/3)\,\phi_{\m\n\r}-3D_{(\m}D.\phi_{\n\r)}+
3D_{(\m}D_{\n}\phi_{\r)\s}{}^\s\nn\\
&-&3g_{(\m\n}\,\Big(
(\DB1-m^2+11\L/3)\,\phi_{\r)\s}{}^\s
-D.D.\phi_{\r)\s}{}^\s
+\frac{1}{2}\,D_{\r)} D.\phi_\s{}^\s\Big)\,\nn\\
&+&\,\frac{3m}{4}\,g_{(\m\n}\,D_{\r)}\chi= 0\,  ,\label{w}\\
{\cal G}_5&=&\frac{3}{2}\,(\DB0-4m^2+8\L)\,\chi+m\,D.\phi_\s{}^\s\, = 0\, .
\label{izard}
\eea
The field $\phi_{\m\n\r}$ is a symmetric 3-tensor and the
auxiliary field $\chi$ decouples at $m=0$ (the strictly massless theory).
Fixing
the ordering of covariant derivatives as shown and requiring 
constraints to remove all but the physical $7=2s+1$ degrees of freedom,
uniquely specifies all terms with an explicit $\L$-dependence.
Indeed, we find the following constraints
\bea
{\cal B}_{\{\n\r\}}
\!\!\!\!&\equiv&D.{\cal G}_{\{\n\r\}}=-\frac{1}{2}\,m\,
\Big(\,D_{\{\n}D_{\r\}}\chi+2m\,D.\phi_{\{\n\r\}}
-4\,m\,D_{\{\n}\phi_{\r\}\s}{}^\s\Big)\, ,\nn\\ \label{B1} \\
{\cal B}_\r\!\!&\equiv&
D.{\cal B}_\r-\frac{m}{4}\,D_\r{\cal G}_5
+\frac{m^4}{4}\,{\cal G}_{\r\s}{}^\s=
\frac{5}{8}\,m\,(3m^2-4\L)\,(D_\r\chi+\frac{2}{3}\,m\,\phi_{\r\s}{}^\s)
\, ,
\nn\\ \\
{\cal B}\!\!&\equiv&
D.{\cal B}-\frac{5}{12}\,m\,(3m^2-4\L)\,{\cal G}_5=
\frac{5}{2}\,m\,(3m^2-4\L)\,(m^2-2\L)\,\chi\, .\nn \\
\label{B3}
\eea
The explicit tensorial structures on the right hand sides 
of~\eqn{B1} are the covariantizations of the flat space ones
in~\eqn{sunk}-\eqn{grunge}.  
The new phenomenon is the splitting of the  prefactors to
$m$, $(3m^2-4\L)$ and $(m^2-2\L)$, thanks to the additional parameter $\L$.
Therefore, in addition to the usual massless theory
at $m=0$ there are new gauge invariant systems at $m^2=4\L/3$ and
$m^2=2\L$, since whenever these prefactors vanish, the corresponding
constraints in~\eqn{B1}-\eqn{B3} become Bianchi
identities with accompanying, respective,  gauge invariances
\bea
\delta \phi_{\m\n\r}=D_{(\m}\,\xi_{\{\n\r\})}\; ,
&&
\delta\chi=0\; ; 
\label{holeyghost}\\
\delta \phi_{\m\n\r}=D_{(\m}D_{\{\n}\,
\xi_{\r\})}+\frac{\L}{3}\,g_{(\m\n}\,\xi_{\r)}\; ,
&&
\delta \chi=-\frac{2}{3}\,\sqrt{\frac{\L}{3}}\,D.\xi
\; ;\label{sun}\\
\delta \phi_{\m\n\r}=D_{(\m}D_{\{\n}D_{\r\})}\,\xi
+\frac{\L}{2}\,g_{(\m\n}D_{\r)}\,\xi\; ,
&&
\delta \chi=-\frac{2}{3}\sqrt{\frac{\L}{2}}\,(D^2+\frac{10\L}{3})\,\xi
\, .\label{fader}\quad
\eea
The new gauge invariant lines
bound regions 
in the $(m^2,\L)$ half-plane
whose unitarity properties are determined by the signs of the
prefactors $(3m^2-4\L)$ and $(m^2-2\L)$. 
To analyze these new properties, decompose the $7=2s+1$
physical DoF into helicities
$(\pm3,\pm2,\pm1,0)$.
We find the following ``phase'' structure of the 
$(m^2,\L)$ half-plane 
\begin{itemize}
\item \underline{$m^2>2\L>0\, :$} This region includes Minkowski 
space and is clearly unitary. All helicities $(\pm3,\pm2,\pm1,0)$,
propagate with positive norm.
\item \underline{$m^2=2\L\, :$} A partially massless theory 
appears since the
scalar constraint ${\cal B}=0$ is now a Bianchi identity whose 
associated gauge invariance removes the scalar helicity
$0$ excitation. The remaining 6 DoF, $(\pm3,\pm2,\pm1)$,
propagate with 
positive norm since they are unaffected by the 
scalar gauge invariance.
\item \underline{$4\L/3<m^2<2\L\, :$} Although all 7 DoF are now
again propagating, the scalar helicity $0$ mode reemerges
{}from the gauge boundary $m^2=2\L$ with negative norm (since the
factor $(3m^2-4\L)\,(m^2-2\L)$ 
appears in canonical commutators as a negative
denominator). This is a unitarily forbidden region.
\item \underline{$m^2=4\L/3\, :$} This partially massless theory has
Bianchi identities ${\cal B}=0={\cal B}_\r$ whose gauge
invariances excise the helicities $(\pm1,0)$.
The remaining 4 PDoF, helicities $(\pm3,\pm2)$
propagate with 
positive norm.
\item \underline{$0<m^2<4\L/3\, :$} Again all 7 DoF propagate but
now the scalar helicity has again positive norm since its denominator
$(3m^2-4\L)\,(m^2-2\L)$ is again positive. However, the region is still
unitarily forbidden because now the vector helicities $\pm1$
suffer a negative denominator $(3m^2-4\L)$. 
\item \underline{$m^2=0\, :$} This is the unitary strictly massless model
with tensor Bianchi identity ${\cal B}_{\{\n\r\}}=0={\cal B}_\r={\cal
B}$.
Only the uppermost helicities $\pm3$ 
remain. An added subtlety is the remnant decoupled  auxiliary field
$\chi$.
\end{itemize}
Notice how the uppermost helicity
$\pm3$    
always emerges unscathed as a pair of positive norm states but,
unlike Minkowski space, splittings into theories with intermediate
lower helicities, rather than only the full complement of $2s+1$
states of the massive theory, are possible. 
Furthermore, the ordering of gauge lines
is the same as for the $s=5/2$ example. But, unlike for 
$s=5/2$, the lines can be
traversed in the order required for unitarity of the partially
massless dS theories without recourse to unphysical, negative, values of $m^2$.

\section{Conclusions}

Life in cosmological backgrounds~\cite{Riess:1998} promises to be less
degenerate than its Minkowski limit. In this paper, we have examined
the notion of mass for higher spins in (A)dS geometries and found a 
far richer structure than in flat space, where higher spin fields are either
strictly massless or massive. Once a cosmological constant is present,
the Minkowski gauge invariances implying masslessness split into
a set of partial ones allowing intermediate, partially massless
theories along lines in the $(m^2,\L)$ plane. 
Their unitarity is easily analyzed by
examining the line ordering.
Although interactions of
relativistic higher spin fields are fraught with many
difficulties\footnote{For massless fields, it has been suggested that
these problems can be alleviated in AdS
backgrounds~\cite{Vasiliev:2000rn}.} even in flat backgrounds, 
and the
observed cosmological scales are far removed from those encountered
in (for example) nuclear physics, this new phenomenon is rather fascinating.
However, many open questions remain, some of which are:

Perhaps the most pressing is an explicit generalization
of our results, beyond simple
counting arguments, to generic values of $s$. 
It is clear that partial masslessness applies to higher values of
$s$ when $m^2$ is appropriately tuned to  $\L$, thereby converting  massive 
constraints to Bianchi identities.
Yet a simple formalism for deducing the pertinent tunings of of $m^2$
would be extremely useful since the unitarity properties of these
partially massless theories can then be deduced by inspection by studying
the ordering of the gauge lines.
Although this problem 
may appear technically challenging, an interesting approach is
suggested by the flat space KK descent methods presented in the
Appendix.

Detailed knowledge of models with $s>3$ would reveal whether the failure of the
$s=5/2$
partially massless theories to be unitary is a universal fermionic
characteristic. 
This question is also related to the lack of
unitary
supergravity theories in dS spaces~\cite{Pilch:1985aw}.
One would also like to know whether auxiliary fields of the 
massless theories decouple, not only in the strictly massless
limit, but also partially massless ones.

Another interesting issue is that of the higher spin version of the $s=2$
van Dam--Veltman--Zhakarov discontinuity for massless limits of 
massive $s=2$ exchange
processes~\cite{vanDam:1970vg}. 
Clearly, if higher spins are allowed to run in
loops, there will generically be discontinuities thanks to the jump in
the virtual DoF over which one functionally integrates~\cite{Dilkes:2001av}. 
A much simpler physical problem, whose eikonal limit is
purely classical, is that of exchange processes mediated by massive
higher spins.
For $s=2$, the flat space discontinuity of the massless limit can be
cured by taking the limit in a $\L\neq0$
background~\cite{Kogan:2000uy}. 
The same
flat space difficulty extends to $s=3/2$~\cite{Deser:1977ur} 
but is again removed
in a cosmological background\footnote{Curiously the limit
$m\rightarrow0$, to a non-unitary massive theory, in AdS is also 
continuous~\cite{Grassi:2001dm}.}~\cite{Deser:2001de}. 
The same question can be posed for
higher $s$ both in strictly and partially massless limits.

An interesting feature of the $m^2=2\L/3$ partially massless $s=2$ 
theory is
that the remaining helicity $(\pm2,\pm1)$ 
excitations propagate on the null cone
of the conformally flat dS metric~\cite{Deser:1983tm}. 
Do partially massless $s>2$ theories also exhibit this conformal behavior?

\begin{appendix}

\section{Appendix: 
Massive Constraints from Dimensional Reduction of Bianchi Identities}

The (A)dS higher spin wave equations presented in this paper were obtained 
by starting with their flat space antecedents, coupling minimally to gravity
and then adding the (unique) non-minimal terms required for constraints
to produce the correct PDoF count. Although both 
massless~\cite{Fronsdal:1978rb,deWit:1980pe}
and massive~\cite{Singh:1974rc} 
higher spin Minkowski field equations are
available, for our purposes a derivation simultaneously yielding
the massive constraints is desirable. Fortunately, dimensional
reduction provides the desired derivation of the massive field
equations from their simpler, $d=5$ massless ancestors~\cite{Aragone:1987yx}.
The key idea is that a massless field 
has the same DoF count as a massive one in one dimension lower.  
In this Appendix, we summarize the dimensional reduction method and
show, in flat space, 
that it allows the massive constraints to be derived from the
Bianchi identities of the massless theories. [It would be
most interesting to rederive the (A)dS results of our paper by
dimensional reduction in a curved background, a problem we reserve for
future investigations.] The methods reported here for $s=5/2$ and
$s=3$, of course,  apply equally well to other values of $s$.

\subsection{Free Massive Spin~5/2}

\label{5/2fl}

The massless $s=5/2$ action and field equations
are
\be
{\cal L}=
-\frac{1}{2}\,\psib^{MN}{\cal R}_{MN}
\label{five}
\ee
\bea
{\cal R}_{MN}
&=&\wh \d\,\psi_{MN}-2\,\G_{(M}\d.\psi_{N)}-2\,\d_{(M}\G.\psi_{N)}
+2\,\G_{(M}\wh \d\,\G.\psi_{N)}\\
&+&\eta_{MN}\,\G.\d.\psi+(\d_{(M}\G_{N)}-\frac{1}{2}\,\eta_{MN}\,\wh
\d\,)\,\psi_R{}^R\,=\,0\ ,
\label{eemi}
\eea
where, in $d=5$,  $M=(\m,5)$ and
Dirac matrices $\G_M$ are defined analogously to our $d=4$
conventions and $\wh X\equiv \G.X$. 
The action enjoys the gauge invariance
\be
\delta \Psi_{MN}=\d_{(M}\ve_{N)}
\label{rail}
\ee
subject to
\be
\G.\ve=0\, ,
\ee
thanks to the Bianchi identity
\be
\d.{\cal R}_N=\frac{1}{5}\,\G_N\,\G.\d.{\cal R}\ .
\label{Blanc}
\ee
Before proceeding, observe that the massless
field equations~\eqn{eemi} in $d=5$ can be rewritten in the simple Christoffel 
form~\cite{deWit:1980pe,Damour:1987vm}
\be
{\cal C}_{MN}\equiv
{\cal R}_{MN}-\frac{1}{5}\,\eta_{MN}{\cal R}_R{}^R
-\frac{2}{5}\,\G_{(M}\G.{\cal R}_{N)}
=\wh\d\,\psi_{MN}-2\,\d_{(M}\G.\psi_{N)}=0\, .
\label{simple}
\ee
We now dimensionally reduce:
First, we would like to gauge  
away components carrying the index ``$5$'', so examine the
gauge transformations
\bea
\delta \psi_{55}&=& \d_5 \ve_5\, ,\\
\delta \psi_{5\m}&=& \frac{1}{2}\,\Big(\d_5\ve_\m+\d_\m\ve_5\Big)\, .
\eea
Assuming that $\d_5$ is invertible, we replace
\be
\d_5\rightarrow im
\ee 
where $m$ will be the $d=4$ mass.
Hence, $\psi_{55}$ may be gauged away.
However the gamma-trace condition on the gauge parameter $\G.\e=0$ prevents
us from doing the same to all components of $\psi_{5\m}$.
[This is the reason~\cite{Aragone:1987yx} auxiliary fields cannot be
avoided beyond $s=2$.]
Instead, we choose a gauge in which only the $(d=4)$ gamma-traceless components
are removed\footnote{Conversely, it may be tempting to gauge away $\psi_{5\m}$
and retain $\psi_{55}$ as an auxiliary field. This gauge choice does
not completely fix the gauge, and the resulting system of equations is
invariant under residual gauge transformations $\delta \psi_{55}=2m\alpha$
and $\delta \psi_{\m\n}=(1/2m)\,\d_\m\d_\n\alpha$ where the gauge
parameter satisfies $(\sl \d+m)\,\alpha=0$. 
One obtains the correct DoF upon fixing this
remaining freedom. In fact there are a
wide range of (residual) gauge invariant equations describing a
massive $s=5/2$ field depending on the details of the gauge choice
made. The choice~\eqn{gauge} has the advantage of completely fixing the
gauge freedom algebraically.}
\be
\psi_{5\m}-\frac{1}{4}\,\g_\m\g.\psi_{5}\equiv\psi_{5\wt\m}=0\, ,
\qquad \psi_{55}=0\, ,
\label{gauge}
\ee
where the $d=4$ Dirac matrices are
\be
\g_\m\equiv i\G_\m\G_5\ ,\qquad\g_5\equiv\G_5\, .
\ee
In terms of
the auxiliary Dirac spinor
\be
\chi\equiv-i\g^\m\psi_{5\m}\;\Longrightarrow\;
\psi_{5\m}=\frac{i}{4}\,
\g_\m\,\chi \ ,
\ee
the equations of motion~\eqn{simple} read
\bea
-i\G_5\,{\cal C}_{\m\n}&=&(\sl\d+m)\psi_{\m\n}-2\,\d_{(\m}
\Big(\g.\psi_{\n)}+\frac{1}{4}\,\g_{\n)}\chi\Big)
\;=\;0\ ,
\label{1}
\\
-i\G_5\,{\cal C}_{5\m}&=&-i\,\Big(m\,\g.\psi_\m
+\frac{1}{2}\,\Big(\d_\m+\frac{1}{2}\,\g_\m
\sl\d\Big)\,\chi\Big)\;=\;0\ ,
\label{2}
\\
-i\G_5\,{\cal C}_{55}&=&2m\chi\;=\;0\, ;
\label{3}
\eea
they imply
\be
(\sl\d+m)\psi_{\m\n}=0\,,\qquad\g.\psi_\m=0\;\Rightarrow\;\d.\psi_\m=0=\psi\, ,
\ee
These are the usual on-shell conditions for a 
massive $d=4$,  $s=5/2$ field. In particular observe that we can identify
the constraints as ${\cal C}_{5\wt\m}$ and ${\cal C}_{55}$.
We stress that the field equations in their simplest form~\eqn{1}-\eqn{3}
do {\it not} directly follow from the variation of an action. 
However the massive $d=4$ action can be obtained from its $d=5$ massless
counterpart by
imposing the gauge condition~\eqn{gauge} and performing the KK descent:
\be
{\cal L}=
-\frac{1}{2}\,\psib^{\m\n}{\cal R}_{\m\n}
-\frac{i}{4}\,\chib\,\g^\m{\cal R}_{5\m}\ .
\label{actnow}
\ee
Here we have relabeled  $-i\G_5\,{\cal R}_{MN}\rightarrow{\cal R}_{MN}$
and $\psib^{\m\n}\equiv\psi^\dagger{}^{\m\n}i\g^0$ is the $d=4$ Dirac conjugate
(remember that $\G^0=-i\g^0\G^5$).
The field equations are
\bea
{\cal R}_{\m\n}&=&\sl \d\,\psi_{\m\n}-2\,\g_{(\m}\d.\psi_\n
-2\,\d_{(\m}\g.\psi_{\n)}+2\,\g_{(\m}\sl \d\,\g.\psi_{\n)}
\nn\\
&+&\eta_{\m\n}\,\g.\d.\psi+(\d_{(\m}\g_{\n)}-\frac{1}{2}\,\sl \d\,)\,
(\psi_\r{}^\r+\frac{1}{2}\,\chi)
\nn\\
&+&m\,(\psi_{\m\n}-2\,\g_{(\m}\g.\psi_{\n)}-\frac{1}{2}\,\eta_{\m\n}\,
\psi_\r{}^\r)\,+\,m\,\eta_{\m\n}\,\chi \,=\,0\; ,\label{redeq1}\\
\frac{i}{4}\g^\m{\cal R}_{5\m}&=&-\,\frac{1}{2}\,(\sl \d-2m)\,\chi
+\frac{1}{4}\,\d.\g.\psi
-\frac{1}{8}\,(\sl \d+4m)\,\psi\, =\,0\; .\label{redeq2}
\eea
After the field redefinition 
\be\psi_{\m\n}\rightarrow\psi_{\m\n}-\frac{1}{8}\,\eta_{\m\n}\,\chi\,
, \label{redefin}\ee
the auxiliary field $\chi$ couples only via $m\,\ol\chi\, \psi_\r{}^\r$ and the
field equations are then the flat limit of the ones presented
in~\eqn{RR} and~\eqn{R5}.

Finally, we show how to deduce the constraints ${\cal C}_{5\wt\m}$ and
${\cal C}_{55}$ from the Bianchi
identity~\eqn{Blanc}: We need  the remaining
components
${\cal R}_{5\wt \m}$ and ${\cal R}_{55}$ of ${\cal R}_{MN}$, 
since the complete set ${\cal R}_{MN}$
inserted in~\eqn{simple} provides the equations ${\cal C}_{MN}$.
The Bianchi identity~\eqn{Blanc} rewritten as
\be
im\,{\cal R}_{5N}=-\d^\m {\cal R}_{\m
N}+\frac{1}{5}\,\G_N\,\d.\G.{\cal R}\ ,
\label{B2} 
\ee
can be used to ``reconstitute'' the missing equations, in particular
\be
{\cal R}_{5\wt\n}=\frac{i}{m}\,\d^\m\,{\cal R}_{\m\wt\n}\, , 
\label{twa}
\ee
and in turn
\be
{\cal R}_{55}=-\frac{1}{m^2}\,\d^\r\d^\s\,{\cal R}_{\r\s}+\frac{1}{4m^2}\,
(\sl\d-m)\,(\d^\r\g^\s\,{\cal R}_{\r\s}+im\g^\r\,{\cal R}_{5\r}) \, .
\label{boac}
\ee
Clearly, since~\eqn{B2} is an {\it identity}, the system of
equations~\eqn{redeq1} and~\eqn{redeq2} 
along with~\eqn{twa} and~\eqn{boac} are
guaranteed to be equivalent to~\eqn{1}-\eqn{3}.
when inserted in the definition~\eqn{simple}.
It is easy to verify that this calculation combined with the
field redefinition~\eqn{redefin} yields the flat limit of the
constraints~\eqn{C51} and~\eqn{c1}.
Incidentally, this proves that the 
action~\eqn{actnow} describes
the correct $2s+1=6$ PDoF.

\subsection{Free Massive Spin~3}

\label{3fl}

The $s=3$, $d=5$  massless Lagrangian and field equations are
\be
{\cal L}=\frac{1}{2}\,\phi_{MNR} {\cal G}^{MNR}
\ ,
\label{lag3}
\ee 
\bea
{\cal G}_{MNR}&=&
\Box_5\,\phi_{MNR}-3\,\d_{(M}\d.\phi_{NR)}+
3\,\d_{(M}\d_{N}\phi_{R)S}{}^S\nn\\
&-&3\,\eta_{(MN}\,\Big(
\Box_5\,\phi_{R)S}{}^S
-\d.\d.\phi_{R)S}{}^S
+\frac{1}{2}\,\d_{R)} \d.\phi_S{}^S\Big)\,= 0 \ .
\eea
The action~\eqn{lag3} has the gauge invariance
\be
\delta\phi_{MNR}=\d_{(M}\xi_{NR)}\, ,
\ee 
with the restriction
\be
\xi_N{}^N=0\, ,
\ee
corresponding to 
the Bianchi identity
\be
\d.{\cal G}_{MN}
=\,\frac{1}{5}\,\eta_{MN}\,\d.{\cal G}_R{}^R\, .
\label{Bianchi3}
\ee 
Again, there is a simpler Christoffel form for the field 
equations~\cite{deWit:1980pe}
\bea
{\cal B}_{MNR}&\equiv&{\cal G}_{MNR}\,-\,\frac{3}{5}\,
\eta_{(MN}{\cal G}_{R)S}{}^S\nn\\
&=&\Box_5\,\phi_{MNR}\,-\,3\,\d_{(M}\d.\phi_{NR)}\,+
\,3\,\d_{(M}\d_N\phi_{R)S}{}^S\, =\,0\ .
\label{sneezy}
\eea
The massive field content is obtained 
by algebraically gauging away the components
\be
\phi_{555}=0\;,\quad \phi_{\m55}=0\;,\quad 
\phi_{5\{\m\n\}}\equiv
\phi_{5\m\n}\,-\,\frac{1}{4}\,\eta_{\m\n}\phi_{5\r}{}^\r\,=\,0 \, .
\ee
There are no residual gauge invariances; the 
remaining fields are $\phi_{\m\n\r}$ 
and the scalar auxiliary
\be
\chi\equiv-i\phi_{5\m}{}^\m\;\Rightarrow\,
\phi_{\m\n5}=\frac{i}{4}\,\eta_{\m\n}\,\chi\, .
\ee
Upon KK descent, $\d_5\rightarrow im$ and the field
equations~\eqn{sneezy} become
\bea
{\cal B}_{\m\n\r}&=&(\Box-m^2)\phi_{\m\n\r}\,
-\,3\,\d_{(\m}\d.\phi_{\n\r)}\,
+\,3\,\d_{(\m}\d_\n\phi_{\r)\s}{}^\s
+\,\frac{3m}{4}\,\eta_{(\m\n}\d_{\r)}\chi\, ,
\nn\\\label{spunk}\\ 
{\cal B}_{\m\n5}&=&\frac{i}{4}\,\eta_{\m\n}\,\Box\,\chi\,
+\,\frac{i}{2}\,\d_\m\d_\n\chi\,
\,-\,im\,(\d.\phi_{\m\n}\,-\,2\,\d_{(\m}\phi_{\n)\s}{}^\s)\ ,\label{sunk}\\
{\cal
B}_{\m55}&=&-\frac{3}{2}\,m\,\d_\m\chi\,-\,m^2\,\phi_{\m\s}{}^\s\ ,
\label{punk}\\
{\cal B}_{555}&=&-3im^2\,\chi\label{grunge}\, ,
\eea
{}from which the usual massive $s=3$ on-shell conditions follow immediately,
\be
(\Box-m^2)\,\phi_{\m\n\r}=0\, ,\quad \chi=0=\phi_\m=\d.\phi_{\m\n}\, .
\ee
Clearly ${\cal B}_{\{\m\n\}5}$, ${\cal B}_{\m55}$ and ${\cal B}_{555}$
are constraints.
The $d=4$ massive 
action principle, guaranteed to yield~\eqn{spunk}-\eqn{grunge}
by virtue of the Bianchi identity, is 
\be
{\cal L}=\frac{1}{2}\,\phi^{\m\n\r} {\cal G}_{\m\n\r}\,
+\,\frac{3i}{8}\,\chi\,{\cal G}_{5\m}{}^\m\, .
\ee
The $d=4$
massive field equations read
\bea
{\cal G}_{\m\n\r}&=&(\Box-m^2)\phi_{\m\n\r}\,
-\,3\,\d_{(\m}\d.\phi_{\n\r)}\,+\,3\,\d_{(\m}\d_\n\phi_{\r)}\
+\,\frac{3m}{4}\,\eta_{(\m\n}\d_{\r)}\chi\, ,\nn\\
&&-3\,\eta_{(\m\n}\,
\Big((\Box-m^2)\phi_{\r)}-\d.\d.\phi_{\r)}+\frac{1}{2}\,\d_\r\d.\phi\Big)
\ ,\label{3eq1}
\\
\frac{3i}{4}\,{\cal G}_5&=&\frac{9}{8}\,(\Box-4\,m^2)\chi
+\frac{3m}{4}\,\d.\phi\ .
\label{3eq2}
\eea
These agree with the flat limit of~\eqn{w} and~\eqn{izard}.
The Bianchi identity~\eqn{Bianchi3} can be rewritten as
\be
im\,{\cal G}_{5MN}=-\d^\r{\cal G}_{\r MN}\,+
\,\frac{1}{5}\,\eta_{MN}\d.{\cal G}_R{}^R
\ee
and yields the ``missing'' field equations
\bea
{\cal G}_{\{\m\n\}5}&=&\frac{i}{m}\,\d^\r{\cal G}_{\r\{\m\n\}}\ ,\\
{\cal G}_{\m55}&=&-\frac{1}{m^2}\,\Big(
\d^\r\d^\s{\cal G}_{\s\{\r\m\}}-\frac{im}{4}\,\d_\m{\cal G}_5
\Big)\ ,\\
{\cal G}_{555}&=&-\frac{i}{m^3}\,\d^\m\d^\n\d^\s{\cal G}_{\s\{\m\n\}}\,
-\frac{1}{4m^2}\,(\Box-m^2)\,{\cal G}_5\,
-\frac{i}{4m}\,\d^\r{\cal G}_{\r\s}{}^\s\ .\quad
\eea
Substituting these in the 
constraints ${\cal B}_{\{\m\n\}5}$, ${\cal B}_{\m55}$ and ${\cal B}_{555}$
following from~\eqn{sneezy} yields the flat limit of the ones given
in~\eqn{B1}-\eqn{B3}.

\end{appendix}

\section*{Acknowledgments}
This work was supported by the 
National Science Foundation under grant PHY99-73935.


\begin{thebibliography}{99} 

\bibitem{Deser:1983tm}
S.~Deser and R.~I.~Nepomechie,
Phys.\ Lett.\ B {\bf 132}, 321 (1983);
Annals Phys.\ {\bf 154}, 396 (1984).

\bibitem{Higuchi:1987py}
A.~Higuchi,
Nucl.\ Phys.\ B {\bf 282}, 397 (1987);
{\it ibid} {\bf 325}, 745 (1989);
J.\ Math.\ Phys.\ {\bf 28}, 1553 (1987).

\bibitem{Deser:2001pe}
S.~Deser and A.~Waldron,
``Gauge invariances and phases of massive higher spins in (A)dS,''
hep-th/0102166.

\bibitem{Johnson:1961vt}
K.~Johnson and E.~C.~Sudarshan,
Annals Phys.\ {\bf 13}, 126 (1961).

\bibitem{Velo:1969bt}
G.~Velo and D.~Zwanziger,
Phys.\ Rev.\ {\bf 186}, 1337 (1969);
A.~Shamaly and A.~Z.~Capri, Ann. Phys. {\bf 74} 503, (1972).
S.~Deser, V.~Pascalutsa and A.~Waldron,
Phys.\ Rev.\  {\bf D62}, 105031 (2000)
[hep-th/0003011].

\bibitem{deWit:1980pe}
B.~de Wit and D.~Z.~Freedman,
Phys.\ Rev.\ D {\bf 21}, 358 (1980).

\bibitem{Lichnerowicz:1961}
A. Lichnerowicz, 
Institut des Hautes \'Etudes
Scientifiques, {\bf 10}, 293 (1961);
Bull. Soc. Math. France, {\bf 92}, 11 (1964). 

\bibitem{Aragone:1987yx}
C.~Aragone, S.~Deser and Z.~Yang,
Annals Phys.\ {\bf 179}, 76 (1987).

\bibitem{Boulware:1972}
D. G. Boulware and S.~Deser,
Phys.\ Rev.\  {\bf D6}, 3368 (1972).

\bibitem{Buchbinder:2000ar}
I.~L.~Buchbinder, D.~M.~Gitman, V.~A.~Krykhtin and V.~D.~Pershin,
Nucl.\ Phys.\  {\bf B584}, 615 (2000)
[hep-th/9910188];
I.~L.~Buchbinder, D.~M.~Gitman and V.~D.~Pershin,
Phys.\ Lett.\ B {\bf 492}, 161 (2000)
[hep-th/0006144].

\bibitem{Deser:2001de}
S.~Deser and A.~Waldron,
Phys.\ Lett.\ B {\bf 501}, 134 (2001)
[hep-th/0012014].

\bibitem{Townsend:1977qa}
P.~K.~Townsend,
Phys.\ Rev.\ D {\bf 15}, 2802 (1977);
S.~Deser and B.~Zumino,
Phys.\ Rev.\ Lett.\  {\bf 38}, 1433 (1977).


\bibitem{Witten:2001}
E. Witten, talk presented {\it Strings 2001}, Tata Institute, Mumbai 2001.

\bibitem{DeWitt:1960fc}
B.~S.~DeWitt and R.~W.~Brehme,
Annals Phys.\ {\bf 9}, 220 (1960);
Phys. Rev.\ {\bf 4}, 317 (1960).

\bibitem{Deser:2002}
S.~Deser and A.~Waldron, 
``Stability of Massive Cosmological Gravitons'',
hep-th/0103255.


\bibitem{Bengtsson:1995vn}
I.~Bengtsson,
J.\ Math.\ Phys.\ {\bf 36}, 5805 (1995)
[gr-qc/9411057].

\bibitem{Abbott:1982ff}
L.~F.~Abbott and S.~Deser,
Nucl.\ Phys.\ B {\bf 195}, 76 (1982).

\bibitem{Pilch:1985aw}
K.~Pilch, P.~van Nieuwenhuizen and M.~F.~Sohnius,
Commun.\ Math.\ Phys.\ {\bf 98}, 105 (1985).

\bibitem{Riess:1998} 
P.~Riess, {\it et al.},
Astron. J. {\bf 116}, 1009 (1998) [astro-ph/9805021]; 
P.~M.~Garnavich, {\it et al.},
Astrophys. J. {\bf 509}, 74 (1998) [astro-ph/9806391];
S.~Perlmutter, {\it et al.}, 
Nature {\bf 391}, 51 (1998) [astro-ph/9712212]. 

\bibitem{Vasiliev:2000rn}
M.~A.~Vasiliev,
``Higher spin symmetries, 
star-product and relativistic equations in AdS  space,''
hep-th/0002183;
L.~Brink, R.~R.~Metsaev and M.~A.~Vasiliev,
Nucl.\ Phys.\ B {\bf 586}, 183 (2000)
[hep-th/0005136].

\bibitem{vanDam:1970vg}
H.~van Dam and M.~Veltman,
Nucl.\ Phys.\  {\bf B22}, 397 (1970);
V. I. Zakharov, JETP Lett. {\bf 12}, 312 (1970);
L. D. Faddeev and A. A. Slavnov, Theor. Math.
Phys. {\bf 3}, 18 (1970); S. K. Wong, Phys. Rev. {\bf D3},
945 (1971).

\bibitem{Dilkes:2001av}
F.~A.~Dilkes, M.~J.~Duff, J.~T.~Liu and H.~Sati,
``Quantum $M \longrightarrow0$ ]
discontinuity for massive gravity with a Lambda term,''
hep-th/0102093.

\bibitem{Kogan:2000uy}
I.~I.~Kogan, S.~Mouslopoulos and A.~Papazoglou,
``The $m\longrightarrow 0$ 
limit for massive graviton in dS(4) and AdS(4): 
How to  circumvent the van Dam-Veltman-Zakharov discontinuity,''
hep-th/0011138;
M.~Porrati,
Phys.\ Lett.\ B {\bf 498}, 92 (2001)
[hep-th/0011152].

\bibitem{Deser:1977ur}
S.~Deser, J.~H.~Kay and K.~S.~Stelle,
Phys.\ Rev.\  {\bf D16}, 2448 (1977).

\bibitem{Grassi:2001dm}
P.~A.~Grassi and P.~van Nieuwenhuizen,
Phys.\ Lett.\ B {\bf 499}, 174 (2001)
[hep-th/0011278].

\bibitem{Fronsdal:1978rb}
C.~Fronsdal,
Phys.\ Rev.\ D {\bf 18}, 3624 (1978);
J.~Fang and C.~Fronsdal,
Phys.\ Rev.\ D {\bf 18}, 3630 (1978);
T.~Curtright,
Phys.\ Lett.\ B {\bf 85}, 219 (1979);

\bibitem{Singh:1974rc}
L.~P.~Singh and C.~R.~Hagen,
Phys.\ Rev.\ D {\bf 9}, 910 (1974);
{\it ibid} 898 (1974).
S.~-J.~Chang, Phys.\ Rev.\ {\bf 161}, 1308 (1967).
F.~A.~Berends, J.~W.~van Holten, P.~van Nieuwenhuizen and B.~de Wit,
Nucl.\ Phys.\ B {\bf 154}, 261 (1979).

\bibitem{Damour:1987vm}
T.~Damour and S.~Deser,
Annales Poincare Phys.\ Theor.\ {\bf 47}, 277 (1987).

\end{thebibliography}
\end{document}